\documentclass[runningheads]{llncs}
\usepackage{graphicx}
\usepackage{url}
\usepackage{amsmath}

\begin{document}
\title{
Accelerating iterative linear equation solver using modified
domain-wall fermion matrix in lattice QCD simulations
}
\titlerunning{Modified domain-wall fermion matrix in lattice QCD}
%
\author{
Wei-Lun Chen\inst{1}\orcidID{0000-0002-8542-9126} \and
Issaku Kanamori\inst{2}\orcidID{0000-0003-4467-1052} \and
Hideo Matsufuru\inst{3}\orcidID{0000-0003-1056-3969} \and
Hartmut Neff\inst{4}
}
\authorrunning{W.-L. Chen et al.}
%
\institute{
Particle and Nuclear Physics,
Graduate Institute for Advanced Studies,
Graduate University for Advanced Studies (SOKENDAI),
Oho 1-1, Tsukuba 305-0801, Japan
\email{wlchen@post.kek.jp}
\and
RIKEN Center for Computational Science
7-1-26 Minatojima-minami-machi, Chuo-ku
Kobe, Hyogo 650-0047, Japan
\email{kanamori-i@riken.jp}
\and
Computing Research Center, High Energy Accelerator
Research Organization (KEK) and 
Accelerator Science Program,  
Graduate University for Advanced Studies (SOKENDAI),
Oho 1-1, Tsukuba 305-0801, Japan
\email{hideo.matsufuru@kek.jp}
\and
Luzernerstrasse 43, 6330, Switzerland
\email{hartmutneff@aol.com}
}
\maketitle              
\begin{abstract}

Lattice simulations of Quantum Chromodynamics (QCD) enable one to
calculate the low-energy properties of the strong interaction among
quarks and gluons based on the first principle.
The most time-consuming part of the numerical simulations of lattice QCD
is typically solving a linear equation for the quark matrix.
In particular, a discretized quark formulation called the domain-wall
fermion operator requires a high numerical cost,
while retaining the lattice version of the chiral symmetry
to good precision.
The domain-wall operator is defined on a five-dimensional (5D)
space extending the four-dimensional (4D) spacetime with an extra
fifth coordinate.
After solving the linear equation in 5D space, the result vector
is projected onto the original 4D space.
There is a variant of the domain-wall operator that improves
the convergence of the 5D linear equation while unchanging the 4D
solution vector.
In this paper, we examine how this variant of the domain-wall operator
accelerates the iterative linear equation solver in practical setups.
We also measure the eigenvalues of the operator and compare the
condition number with the convergence of the solver.
We use a generic lattice QCD code set Bridge++ that is planned to
be released including the improved form of the domain-wall operator
examined in this work with code for the GPU.

\end{abstract}

\section{Introduction}
\label{sec:Introduction}

Large-scale simulations of lattice Quantum Chromodynamics (QCD)
have been a challenge for the high-performance computing
for a long time.
The QCD is the fundamental theory of the strong interaction
among quarks and gluons, which is formulated based on the invariance
under the local SU(3) gauge transformation on the color degree
of freedom of the quark and gluon fields.
While the fundamental theory is known, it is quite difficult to
solve it analytically because of the strong coupling of this theory.
The lattice QCD is a field theory formulated on the Euclidean
discretized spacetime.
Employing the path integral quantization, this theory enables
numerical evaluation of the physical quantities using
the Monte Carlo algorithms.
Thus, the lattice QCD simulations provide a general procedure
to investigate QCD based on the first principle.
Such a calculation is particularly important in the search for
phenomena beyond the standard model through precision evaluation
of the hadronic scattering processes.
Lattice QCD is also important in the quantitative understanding
of the nuclear force among hadrons, the phase structure of QCD
at finite temperature and density, and so on.

The most time-consuming part of the lattice QCD simulation is solving
a linear equation for the quark (fermion) matrix, which is typically
large-scale, sparse, and solved by iterative methods.
There is a variety in the form of the fermion matrix since the
only required condition is that it approaches QCD in the continuum
limit.
In this paper, we focus on the type of matrix called
the domain-wall fermion operator.
This fermion matrix has an important feature that preserves the
chiral symmetry, an important symmetry of QCD, on the lattice with
good precision.
The domain-wall operator is defined in the five-dimensional (5D) space,
which extends the four-dimensional (4D) spacetime by adding
the fifth coordinate.
The physical modes in the original four-dimensional space are picked
up from the two edges in the fifth direction.
Since the domain-wall fermion matrix is defined in the five-dimensional
space, the numerical cost is higher than that of other formulations.
Thus improving the convergence of the fermion matrix solver is
a quite important subject in the numerical simulations.

One of the authors (H. Neff) proposed a form of domain-wall operator
that improves the convergence of the solver while unchanging
the physical modes in the four-dimensional space \cite{Neff:2015wsa}.
This paper aims at a systematic and practical investigation
of Neff's form and how it improves the convergence of the solver.
While Ref.~\cite{Neff:2015wsa} already examined the effect of
the improved form on several parameter sets, we add several new setups
including the operator with the link smearing.
We determine the condition number of the matrix by measuring the
lowest and highest eigenvalues of the Hermitian domain-wall matrix
and compare with the convergence of the Conjugate Gradient (CG) solver.
In the numerical study, we employ the Bridge++ code set
\cite{Akahoshi:2021gvk}
with an extension to incorporate Neff's improved form of
the domain-wall matrix.
We use the code for GPU implemented with OpenACC \cite{Chen:2025hpca}.

This paper is organized as follows.
The next section introduces the lattice QCD and our target fermion
matrix in a certain depth.
While the equations necessary to implement the fermion matrix
used in this work are displayed to make the description self-contained,
the essential ingredient is the introduction of a tunable parameter
$\alpha$ in the 5D matrix that does not change the 4D solution vector.
Section~\ref{sec:Results} shows our numerical results.
We examine the effect of $\alpha$ in the modified matrix on
the convergence of CG solver.
The last section is devoted to our conclusion and outlook.

\section{Lattice QCD}
\label{sec:Lattice_QCD}

\subsection{Lattice QCD and chiral symmetry}

The lattice QCD action in four-dimensional Euclidean space is written
as follows:
\begin{equation}
 S_{\rm QCD} = \sum_x \left\{
 (S_G[U_\mu(x)] + \bar{\psi}(x) D_F[U] \psi(x)
 \right\},
 \label{eq:lattice_QCD_action}
\end{equation}
where $S_G$ is a gauge action, $\psi(x)$ and $\bar{\psi}(x)$
anti-commuting Grassmann fields representing quark and anti-quark,
$D_F[U]$ a fermion operator depending on the link variable $U_\mu(x)$,
$x=(x_1, x_2, x_3, x_4)$ a lattice site, where the lattice spacing
$a$ is set to unity for simplicity.
$U_\mu(x)$ is a $3\times 3$ complex matrix field representing
the gauge field.
$\psi$ has the color and spinor degrees of freedom in addition to
the lattice site $x$.
The spinor has 4 components that represent the up and down spin
for quark and anti-quark.
Thus the quark field has $3\times 4=12$ complex components
on each site $x$. 

Employing the path integral quantization, the QCD partition function
reads
\begin{eqnarray}
&&\lefteqn{\int {\cal D}U{\cal D} \psi {\cal D} \bar{\psi}
\exp\left[ -S_G - \sum_x \bar{\psi}(x) D_F[U] \psi(x) \right]
= \int {\cal D} U \det (D_F) e^{-S_G}}
\nonumber\\
&& \hspace{1cm}
= \int {\cal D}U{\cal D} \phi {\cal D} \phi^\dagger \exp\left[
    -S_G -\frac{1}{2} \sum_x\phi^\dagger(x) D_F[U]^{-1} \phi(x) \right],
\end{eqnarray}
where complex (ordinary number) field $\phi(x)$ is introduced as
Gaussian integration variables.
Applying a Monte Carlo method, the expectation value of a physical observable
${\cal O}$ is represented as
\begin{equation}
 \langle {\cal O}[U,\psi, \bar{\psi}] \rangle
 = \frac{1}{N}\sum_i^N O[U^{(i)}, S_q^{(i)}].
\end{equation}
Here, the gauge field configurations $\{U^{(i)}\}$ are assumed to be
generated with the Boltzmann weight 
$\exp[-S_G]\det(D_F)$, and $S_q^{(i)}$ is a quark propagator
obtained by solving a linear equation for the fermion matrix $D_F$,
\begin{equation}
 D_F[U^{(i)}]_{y,x} S_q^{(i)}(x) = b_y,
\label{eq:linear_equation}
\end{equation}
where the color and spinor indices are omitted for simplicity.
To obtain the physical quantities with high precision,
one needs to solve this linear equation
many times
to calculate $O[U^{(i)}, S_q^{(i)}]$ as well as during the generation of
the gauge configurations.
Thus solving this linear equation is a typical bottleneck of lattice
QCD simulations.

In discretizing the continuum QCD action,
there is a variation of the action on the lattice,
since the only required condition is that the lattice
action approaches to the QCD action in the continuum limit,
$a\rightarrow 0$, where $a$ is the lattice spacing.
Thus a number of lattice actions have been proposed
intending a rapid approach to the continuum limit.
Each lattice fermion action has its own pros and cons.
One of the most popular fermion operators is the Wilson fermion,
\begin{eqnarray}
  D_W(x,y; M) 
  &=& (4+M) \delta_{x,y}
   - \frac{1}{2}\sum_{\mu=1}^4 \big\{ \,
      (1-\gamma_\mu) U_\mu(x)\delta_{x+\hat{\mu},y}
      \nonumber\\
  & & \hspace{3.5cm} 
  + (1+\gamma_\mu) U_\mu^\dagger (x-\hat{\mu})
                           \delta_{x-\hat{\mu},y}  \big\} ,
\label{eq:Wilson_matrix}
\end{eqnarray}
where $M$ is quark mass, $\hat{\mu}$ is a unit vector in
$\mu$-th direction ($\mu=1,2,3,4$), $\gamma_\mu$ is a
$4\times 4$ complex matrix acting on the spinor components.
The improved form of the Wilson fermion action, called
clover fermion, is extensively used in large-scale simulations
of lattice QCD.
It has a disadvantage, however, that a symmetry called
chiral symmetry, which is satisfied by the QCD action for
vanishing quark mass, is explicitly violated even with
setting $M=0$.

In the continuum QCD, the chiral symmetry is represented as
\begin{equation}
    \gamma_5 D_F + D_F \gamma_5 = 0,
\end{equation}
where $\gamma_5 = \gamma_1 \gamma_2 \gamma_3 \gamma_4$.
This symmetry concerns the left- and right-handed components
of the quark field.
The spontaneous breakdown of this symmetry at low-energy
ensures the smallness of the pion masses and 
is attributed to most of the masses of the proton and the neutron.
The chiral symmetry is also the basis of the phenomenological
properties of the hadrons.
Thus the quark action that respects the chiral symmetry has
particular importance in understanding the chiral dynamics of QCD.
For the Wilson fermion action, the chiral symmetry is restored
only in the continuum limit.

On the lattice, the chiral symmetry had been a difficulty
for a long time.
Understanding of the chiral symmetry on the lattice
was proceeded by the discovery of the Ginsparg-Wilson relation \cite{Ginsparg:1981bj},
\begin{equation}
    \gamma_5 D_F + D_F \gamma_5 = a D_F R \gamma_5 D_F,
\end{equation}
where $R$ is a certain constant.
\cite{Hasenfratz:1998ri,Luscher:1998pqa}.
After the realization of the Ginsparg-Wilson relation as the exact
chiral symmetry on the lattice, several fermion actions 
that satisfies the Ginsparg-Wilson relation exactly or
approximately have been proposed.
The Domain-wall fermion is the latter kind of formulation
and in the limit of the infinite extent of the fifth direction,
$L_s\rightarrow \infty$, it satisfies the Ginsparg-Wilson relation
exactly.

\subsection{Domain-wall fermion matrix}

The domain-wall fermion was proposed by Kaplan \cite{Kaplan:1992bt} intending
to formulate the chiral gauge theory.
The domain-wall fermion is defined on the five-dimensional space by adding
fifth coordinate to the four-dimensional spacetime, and a large negative mass
is introduced to the Wilson operator kernel (\ref{eq:Wilson_matrix}).
The light modes appear on the two edges in the fifth direction.
After the realization of the domain-wall fermion as a light fermion formulation
\cite{Shamir:1993zy,Furman:1994ky}, several improved forms have been
proposed \cite{Borici:1999da,Brower:2004xi,Chiu:2002ir}.
These variants are described by the following form \cite{Brower:2004xi}:
{\small
\begin{equation}
  D_{DW} = \left(
 \begin{array}{cccccc}
  D_+^{(1)}    & D_-^{(1)} P_-&        0    &   \cdots   & 0 & -mD_-^{(1)}P_+\\
  D_-^{(2)} P_+& D_+^{(2)}    & D_-^{(2)} P_-&            &   &      0    \\
   0          & D_-^{(3)} P_+& D_+^{(3)}    & D_-^{(3)} P_-&  &      0    \\
   \vdots  &        & \ddots&        \ddots &     \ddots &  \vdots      \\
    0      &        &       & D_-^{(L_s-1)} P_+& D_+^{(L_s-1)} & D_-^{(L_s-1)} P_-\\
  -mD_-^{(L_s)}P_-&  0 & \cdots&     0         & D_-^{(L_s)}P_+& D_+^{(L_s)}    \\
 \end{array}
\right)
\label{eq:domainwall_general}
\end{equation}
}
where 
 $m$ is quark mass, $P_- = (1-\gamma_5)/2$, $P_+ = (1+\gamma_5)/2$,
\begin{equation}
  D_+^{(i)} = b_i D_W + 1, \hspace{1cm}
  D_-^{(i)} = c_i D_W - 1,
\end{equation}
$D_W=D_W(-M_0)$ is the Wilson operator (\ref{eq:Wilson_matrix})
with large negative mass $-M_0$, where $M_0=O(1)$ is called
the domain-wall height.
There are several choices for the parameters $b_i$ and $c_i$:
$(b_i,c_i)=(1,0)$ \cite{Shamir:1993zy,Furman:1994ky},
$(b_i,c_i)=(1,1)$ \cite{Borici:1999da},
$b_i - c_i = \mbox{const.}$ (independent of $i$) 
called the M\"obius fermion \cite{Brower:2004xi},
and $b_i=c_i$ chosen depending on $i$ so as to optimize the
Ginsparg-Wilson relation \cite{Chiu:2002ir}.
In this paper we only examine the cases $(b_i,c_i)$ independent
of $i$.

In the fifth direction, the following boundary condition
is imposed at $s=1$ and $L_s$:
\begin{equation}
P_+ \psi(s=0) = P_- \psi(s=L_s+1) = 0 .
\end{equation}
The fermion field in four-dimensional space is given as%
\footnote{
This definition of $\bar{q}(x)$ corresponds to $\tilde{q}_x$
called ``traditional choice'' in Ref.~\cite{Brower:2012vk}.
}
\begin{eqnarray}
 q(x) &=& P_- \, \psi(x,s\!=\!1) + P_+ \, \psi(x,s\!=\!L_s),
\nonumber \\
 \bar{q}(x) &=& \bar{\psi}(x,s\!=\!1)(-D_-^{(1)})P_+
                  + \bar{\psi}(x,s\!=\!L_s) (-D_-^{(L_s)}) P_- .
\label{eq:4d_quark_field}
\end{eqnarray}
In the limit of $L_s\rightarrow\infty$, the lattice chiral
symmetry is satisfied exactly.
This limit corresponds to the overlap fermion operator,
$D_{\rm ov}=1+\gamma_5 \mbox{sign}(\gamma_5 D_W)$
in the massless case, which practically requires
large numerical resources.

\paragraph{Link smearing}
Currently the link smearing is frequently adopted as an improvement
procedure for fermion actions.
The link smearing is essentially the integration of the surrounding gauge
field into the link variable, $U_\mu(x)$.
We consider the stout projection combined with the APE smearing
\cite{Morningstar:2003gk}:
\begin{equation}
  C_\mu(x) = \sum_{\nu\neq\mu} \rho \left[
       U_\nu(x)U_\mu(x\!+\!\hat{\nu})U_\nu^\dagger(x\!+\!\hat{\mu})
      + U_\nu^\dagger(x\!-\!\hat{\nu}) U_\mu(x\!-\!\hat{\nu})
         U_\nu(x\!-\!\hat{\nu}\!+\!\hat{\mu}) \right] ,
\nonumber
\end{equation}
\begin{equation}
U_\mu(x)'
= \exp\left( [C_\mu(x)U_\mu^\dagger(x)]_{\rm AT} \right) U_\mu(x),
\label{eq:stout_smearing}
\end{equation}
where $[\cdots]_{\rm AT}$ represents anti-Hermitian and traceless operation,
and $\rho$ is a tunable parameter.
This smearing step (\ref{eq:stout_smearing}) can be repeatedly applied.
The fermion operator can be improved by adopting the smeared
link variable instead of the original one.

\subsection{Better conditioned form of domain-Wall operator}

Improved form of the domain-wall fermion operator which unchanges
the four-dimensional fermion field was proposed by H.~Neff
\cite{Neff:2015wsa}.
This form is represented as follows.
{\small
\begin{equation}
  D_{DW}^{(\alpha)} = \left(
 \begin{array}{cccccc}
  D_+^{(1)}(P_- + \alpha P_+)\ & \alpha D_-^{(1)} P_- &        0    &   \cdots   & -mD_-^{(1)}P_+\\
  \alpha D_-^{(2)} P_+        & \alpha D_+^{(2)}     & \alpha D_-^{(2)} P_-&            &        0    \\
   \vdots        & \ddots&        \ddots &     \ddots &  \vdots      \\
  -mD_-^{(L_s)}P_-&  0 & \cdots&      \alpha D_-^{(L_s)} P_+\ & D_+^{(L_s)} (P_+ + \alpha P_-)    \\
 \end{array}
 \right)
 \label{eq:domainwall_neff}
\end{equation}
}
Note that except for the four corners of the fifth coordinate matrix,
each four-dimensional block is multiplied by $\alpha$.
The top-left and bottom-right blocks are modified by multiplying by
$(P_\mp + \alpha P_\pm)$, respectively, and the top-right and bottom-left
blocks are unchanged.
This modification does not change the four-dimensional quark field,
while it affects the convergence of the linear equation in five-dimensional space.

There is a relation between $D_{DW}^{(\alpha)}$ and $D_{DW}=D_{DW}^{(\alpha=1)}$
that 
\begin{equation}
    D_{DW}^{(\alpha)}\, {\cal P} = D_{DW}^{(\alpha=1)}\, {\cal P}{\cal A} , 
\label{eq:relation_to_alpha=1}
\end{equation}
where
\begin{equation}
  {\cal P} = \left(
 \begin{array}{cccccc}
  P_-    &  P_+  &    0   & \cdots & 0 \\
  0      &  P_-  &   P_+  & \ddots &\vdots   \\ 
  \vdots & \ddots& \ddots & \ddots & 0  \\
  0      &       & \ddots &   P_-  & P_+  \\
  P_+    &   0   & \cdots &    0   & P_- \\
 \end{array},
 \right)
\hspace{0.7cm}
  {\cal A} = \left(
 \begin{array}{cccccc}
  1      &   0    &        & \cdots &  0     \\
  0      & \alpha &   0    &        & \vdots \\ 
  \vdots & \ddots & \ddots & \ddots & \vdots \\
  \vdots &        & \ddots & \alpha &  0     \\
  0      & \cdots & \cdots &   0    & \alpha \\
 \end{array}
 \right) .
\end{equation}
We need to solve a 5D equation
\begin{equation}
  D_{DW}^{(\alpha=1)} x_5 = b_5 ,
\label{eq:5D_linear_equation}
\end{equation}
where according to Eq.~(\ref{eq:4d_quark_field})
$b_5 = (-D_-){\cal P}^\dagger (0, \cdots, 0, b_4)^T$ for a 4D
source vector $b_4$, with
$D_-=\mbox{diag}(D_-^{(1)}, \cdots, D_-^{(L_s)})$.
Once a 5D solution vector $x_5$ is determined, the 4D solution
vector $x_4$ is provided as $x_4= [{\cal P}^\dagger x_5]_{s=1}$.
Due to the relation (\ref{eq:relation_to_alpha=1}),
the solution of $D_{DW}^{(\alpha)}x_5^{(\alpha)}=b_5$
give the same 4D solution $x_4$, since
\begin{equation}
    {\cal P}^\dagger x_5
    = (D_{DW} {\cal P})^{-1} b_5
    =  (D_{DW}^{(\alpha)} {\cal P}{\cal A}^{-1})^{-1} b_5
    = {\cal A} {\cal P}^\dagger (D_{DW}^{(\alpha)})^{-1} b_5
    = {\cal A} {\cal P}^\dagger x_5^{(\alpha)},
\end{equation}
where ${\cal P}^\dagger = {\cal P}^{-1}$,
and $[{\cal A} {\cal P}^\dagger x_5^{(\alpha)}]_{s=1}
= [{\cal P}^\dagger x_5^{(\alpha)}]_{s=1}$.
Thus if the linear equation with $\alpha\neq 1$ can be solved
faster than the original equation (\ref{eq:5D_linear_equation}),
the numerical cost would be reduced.

\paragraph{Even-odd preconditioning}
Noting that the $D_W$ and $D_{DW}$ contain nearest neighbor coupling,
by dividing sites into even and odd sites,
$D_{DW} x=b$ is represented as
\begin{equation}
  D_{DW} \, x = \left( \begin{array}{cc}
          D_{ee} & D_{eo} \\
          D_{oe} & D_{oo} \\
       \end{array} \right)
  \left( \begin{array}{c}
          x_{e} \\
          x_{o} \\
       \end{array} \right)
 = 
  \left( \begin{array}{c}
          b_{e} \\
          b_{o} \\
       \end{array} \right) = b.
\label{eq:eo_decomposition}
\end{equation}
One arrives at the even-odd preconditioned linear equation,
\begin{align}
 D \, x_e
  &\equiv
  (1 - D_{ee}^{-1} D_{eo} D_{oo}^{-1} D_{oe})\, x_e = \tilde{b}_e,
\label{eq:even-odd_linear_equation} \\
\tilde{b}_e &= D_{ee}^{-1} (b_e - D_{eo} D_{oo}^{-1} b_o) ,
\\
x_o &= D_{oo}^{-1} (b_o - D_{oe} x_e) .
\end{align}
One needs to solve Eq.~(\ref{eq:even-odd_linear_equation})
whose matrix $D$ usually has a smaller condition number than $D_{DW}$.
There are two ways to divide the sites into even and odd:
in the five-dimensional space or the four-dimensional space.
We adopt the latter for simplicity and similarity with other
fermion formulations.
In this paper we focus on the effect of $\alpha$ on the
convergence of the iterative solver, and do not concern the details
of implementation and its performance which is described in
\cite{Chen:2025hpca}.

Note that there is another version of even-odd preconditioned
equation,
\begin{equation}
  (1 - D_{eo} D_{oo}^{-1} D_{oe} D_{ee}^{-1})\, y_e = b'_e,
\label{eq:another_even-odd_linear_equation}
\end{equation}
where $b'_e = b_e - D_{eo} D_{oo}^{-1} b_o$ and
$x_e = D_{ee}^{-1} y_e$.
However, the relation (\ref{eq:relation_to_alpha=1}) means
$D_{DW}^{(\alpha)}=D_{DW}({\cal P}{\cal A}^{-1}{\cal P}^\dagger)$,
and thus $D_{ee}^{(\alpha)}=D_{ee}^{(\alpha=1)}({\cal P}{\cal A}^{-1}{\cal P}^\dagger)$
and so on.
In Eq.~(\ref{eq:another_even-odd_linear_equation}), 
$({\cal P}{\cal A}^{-1}{\cal P}^\dagger)$ and
$({\cal P}{\cal A}^{-1}{\cal P}^\dagger)^{-1}={\cal P}{\cal A}{\cal P}^\dagger$
cancel each other, which results in no effect on the convergence
of the solver.

In lattice QCD simulations, the BiCGStab algorithm is often a good choice.
However, for the domain-wall fermion matrix, BiCGStab
does not converge.
This is because the eigenvalues of $D_{DW}$ distribute
in the negative real part.
Thus the CG algorithm for $D^\dagger D$ is applied, where
``$\dagger$'' denotes Hermitian conjugate, {\it i.e.},
taking the complex conjugate and the transpose of the matrix $D$
in Eq.~(\ref{eq:even-odd_linear_equation}).

It is well-known that a single precision solver can work as
a preconditioner in the solver in double precision
which is the precision practically required in numerical
simulations.
In such a multi-precision solver, most of the time is consumed
in the single precision solver, and hence the performance
of matrix multiplication to a vector in single precision
determines the efficiency of solver algorithms.
It is also expected that the convergence property of the
iterative solver does not differ much for double and single
precision.
Thus in the following we show the convergence of the solver
for the single precision solver only.

\section{Results}
\label{sec:Results}

\subsection{Numerical setup}

We perform a numerical experiment to investigate how the introduction
of the parameter $\alpha$ improves the convergence of five-dimensional
solver.
For practical investigation, we generate three ensembles of gauge
configuration generated with quenched approximation, namely,
without the quark vacuum polarization effect.
They are generated with the plaquette gauge action at the gauge
coupling $\beta=6.0$ and 5.7, which roughly correspond to the values
of lattice spacing $a\simeq 0.1$ and 0.2 fm, respectively.
For $\beta=6.0$, two lattice volumes $16^4$ and $32^4$ are adopted.
For the $\beta=5.7$, we generate a $16^4$ lattice that roughly
corresponds to the same physical volume as the $32^4$ lattice
with $\beta=6.0$.
On these configurations, the parameter of the domain-wall operator
$M_0 = 1.8$ is used as a typical value in practical simulations,
as adopted by the RBC/UKQCD Collaboration
\cite{RBCUK:2013}.

To examine the effect of the link smearing, we apply three steps
of the stout projection \cite{Morningstar:2003gk} with the APE
smearing with $\rho = 0.1$ to the $\beta=6.0$ configurations.
This corresponds to the setup adopted by the JLQCD collaboration
\cite{JLQCD:2022}.
In this case, $M_0=1.0$ is adopted since the lattice artifact to shift
the best value of $M_0$ from 1 due to the interaction is largely suppressed.

On these ensembles, the parameters $(b, c) = (1.5, 0.5)$ and
$(1.0, 1.0)$ are adopted with two values of $L_s=8$ and 16.
The values of parameters $(b, c) = (1.5, 0.5)$  are adopted by
JLQCD Collaboration \cite{JLQCD:2022}
and RBC/UKQCD Collaboration \cite{RBCUK:2013}, and
$(b, c) = (1.0, 1.0)$ corresponds to the Borici's form
\cite{Borici:1999da}.
We observe the convergence of the CG solver for several values of
quark mass in the range of $m=$0.001--0.01, with varying
the value of $\alpha$.

As the simulation code, we employ the Bridge++ code set
\cite{Bridge,Akahoshi:2021gvk} 
with an extension to offloading code for the GPU implemented
using OpenACC \cite{Chen:2025hpca}.
The numerical computation is done on the Pegasus system at
University of Tsukuba, whose each node is composed of Intel
Xeon processor and a single NVIDIA H100 GPU.
All the computation is performed on a single process with
single GPU device.
Since this work concerns the algorithmic convergence of an
iterative solver, we do not discuss the performance of
the code which is described in \cite{Chen:2025hpca}.

\subsection{Result for $\beta$ = 6.0 on the $16^4$ lattice}

Let us start with examining the effect of $\alpha$ on the convergence
of CG solver on the $16^4$ lattice generated with the gauge coupling
$\beta =6.0$.
In the unsmeared cases, we set $M_0 = 1.8$ as aforementioned.
The CGNR solver is applied in single precision with
the convergence criterion $|r|^2 < 10^{-14}$.

\begin{figure}[tb]
\centering
\includegraphics[width=0.49\textwidth]{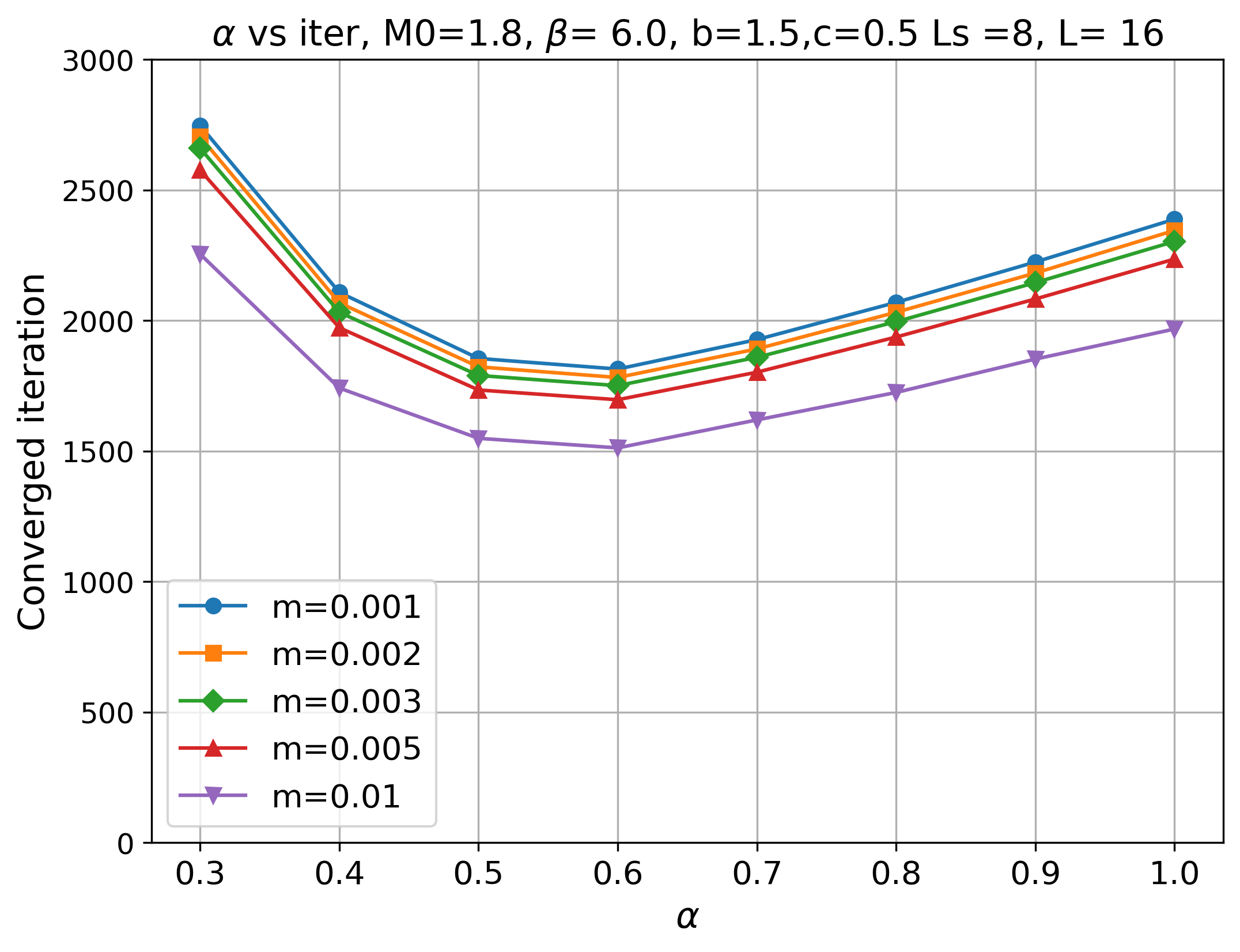}
\includegraphics[width=0.49\textwidth]{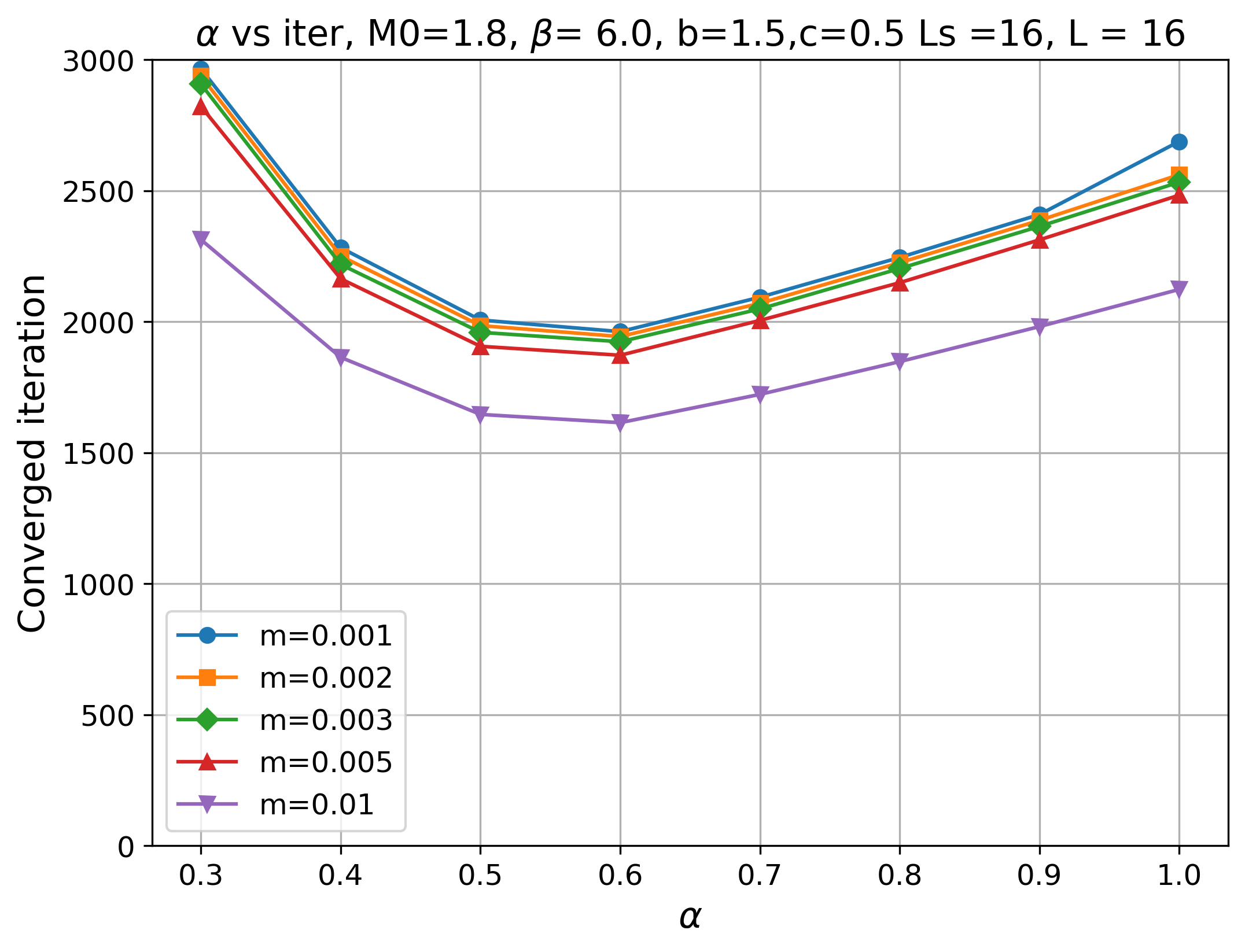}
\includegraphics[width=0.49\textwidth]{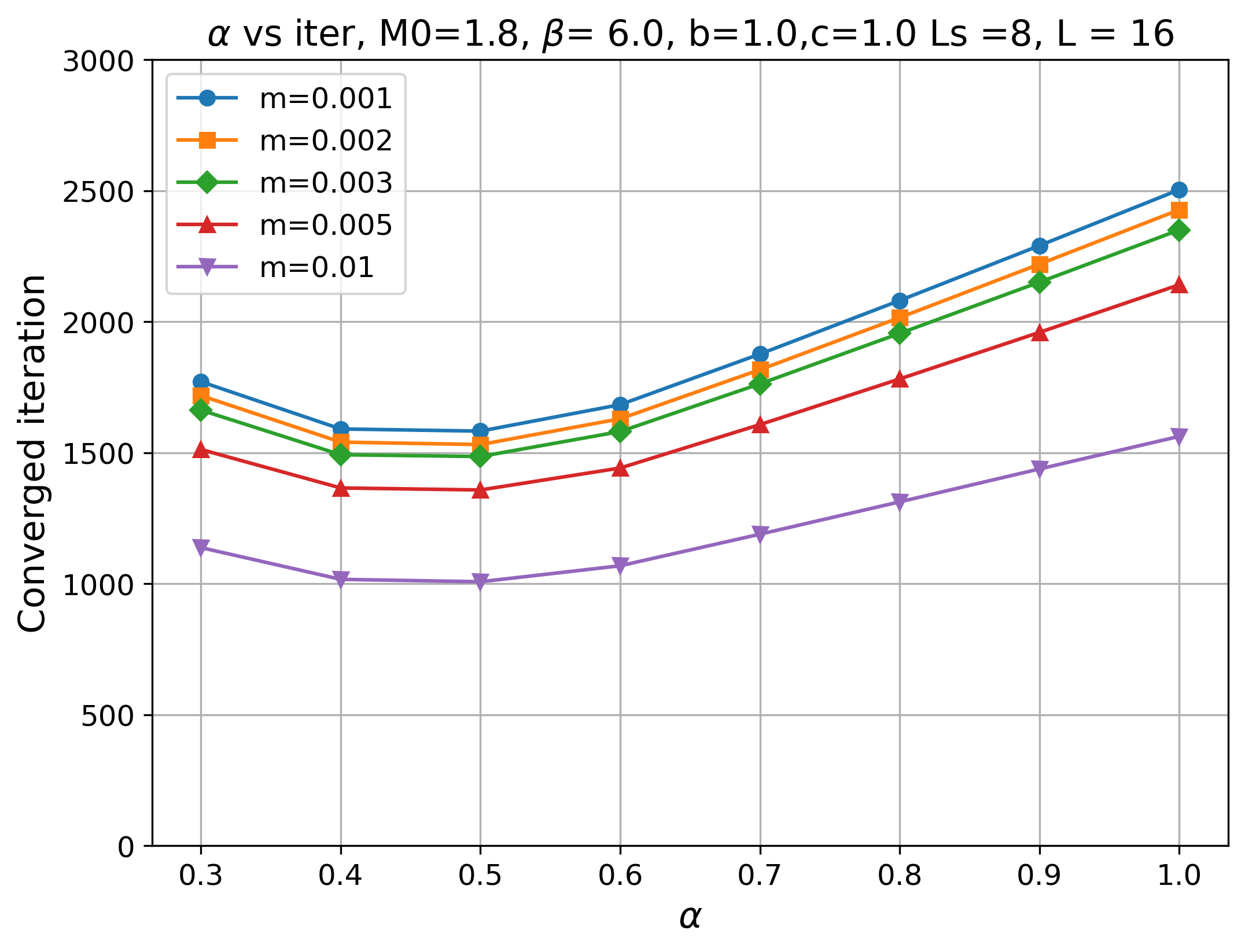}
\includegraphics[width=0.49\textwidth]{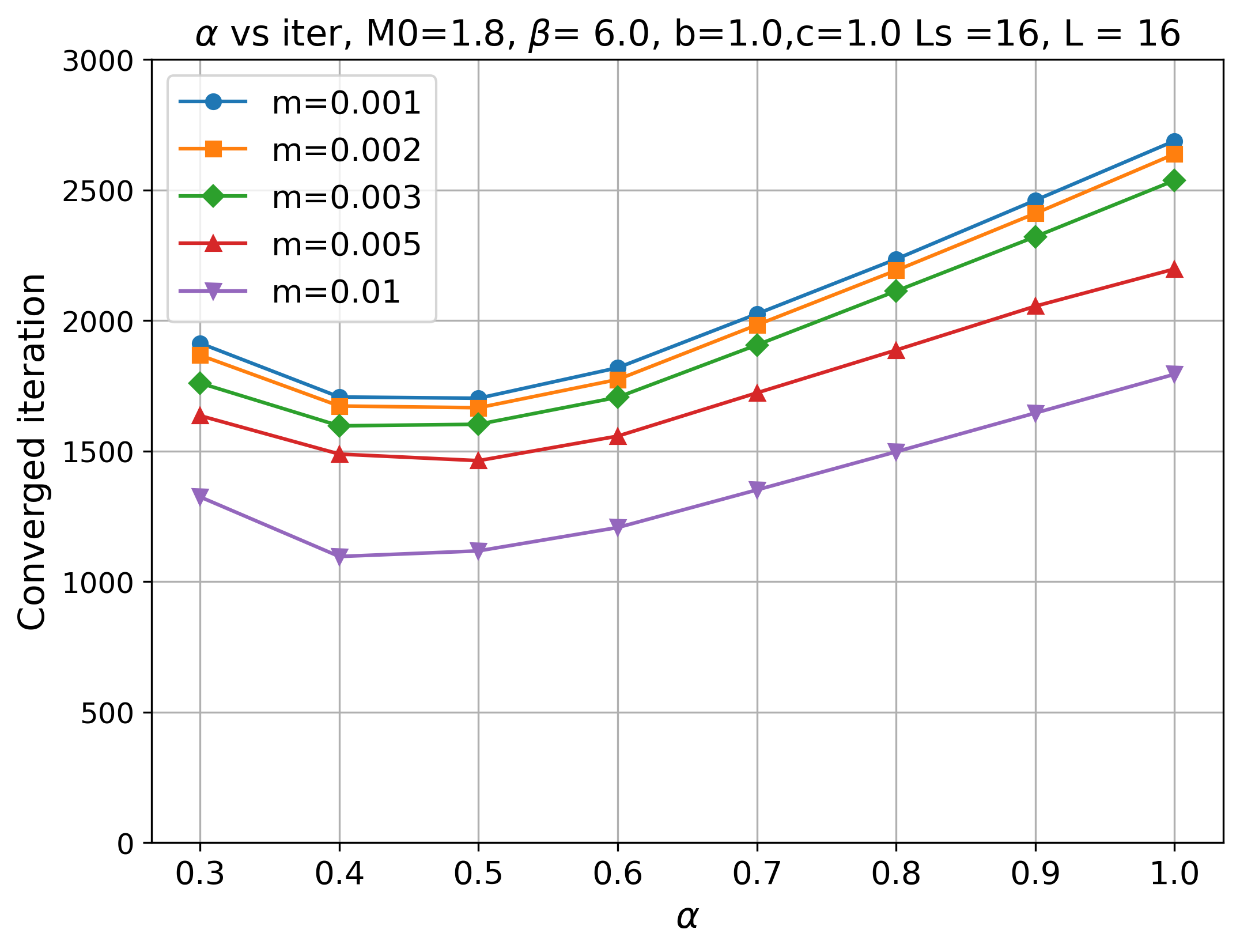}
\caption{
Converged number of CG iteration for $\beta=6.0$ on $(L=16)^4$ lattice,
$M_0 = 1.8$ (without smearing), $(b,c)=(1.5,0.5)$
(top panels) $L_s=8$ (left) and 16 (right)
and $(b,c)=(1.0,1.0)$ (bottom panels).
}
\label{fig:conv_L16_beta6.0_M01.8}
\end{figure}

Figure~\ref{fig:conv_L16_beta6.0_M01.8} shows 
the convergence iteration number against $\alpha$ on 
a single configuration without the link smearing.
Fluctuation against the gauge configuration will be discussed
later.
The top panels of Fig.~\ref{fig:conv_L16_beta6.0_M01.8} 
display the result for $(b,c)=(1.5,0.5)$ with $L_s=8$ and 16.
For $L_s=8$, $\alpha\simeq 0.6$ most improves the convergence,
and 25\% improvement compared to $\alpha=1$ independently of
the quark mass.
While $L_s=16$ requires slightly more iterations until convergence,
the result shows the same tendency that $\alpha\simeq 0.5$
is the best, while the improvement decreases as the quark mass
from 25\% for $m=0.001$ to 20\% for $m=0.01$.

The bottom panels of Figure~\ref{fig:conv_L16_beta6.0_M01.8} 
display the result for $b=1.0$, $c=1.0$ (Borici's setting).
For both the $L_s=8$ and $16$, $\alpha\simeq 0.5$ improves
the convergence most, for which about 36\% acceleration is
achieved independently of the quark mass and $L_s$.
As an observation, the quark mass dependence of converged
iteration is larger than the case of $(b,c)=(1.5,0.5)$.

\begin{figure}[tb]
\centering
\includegraphics[width=0.49\textwidth]{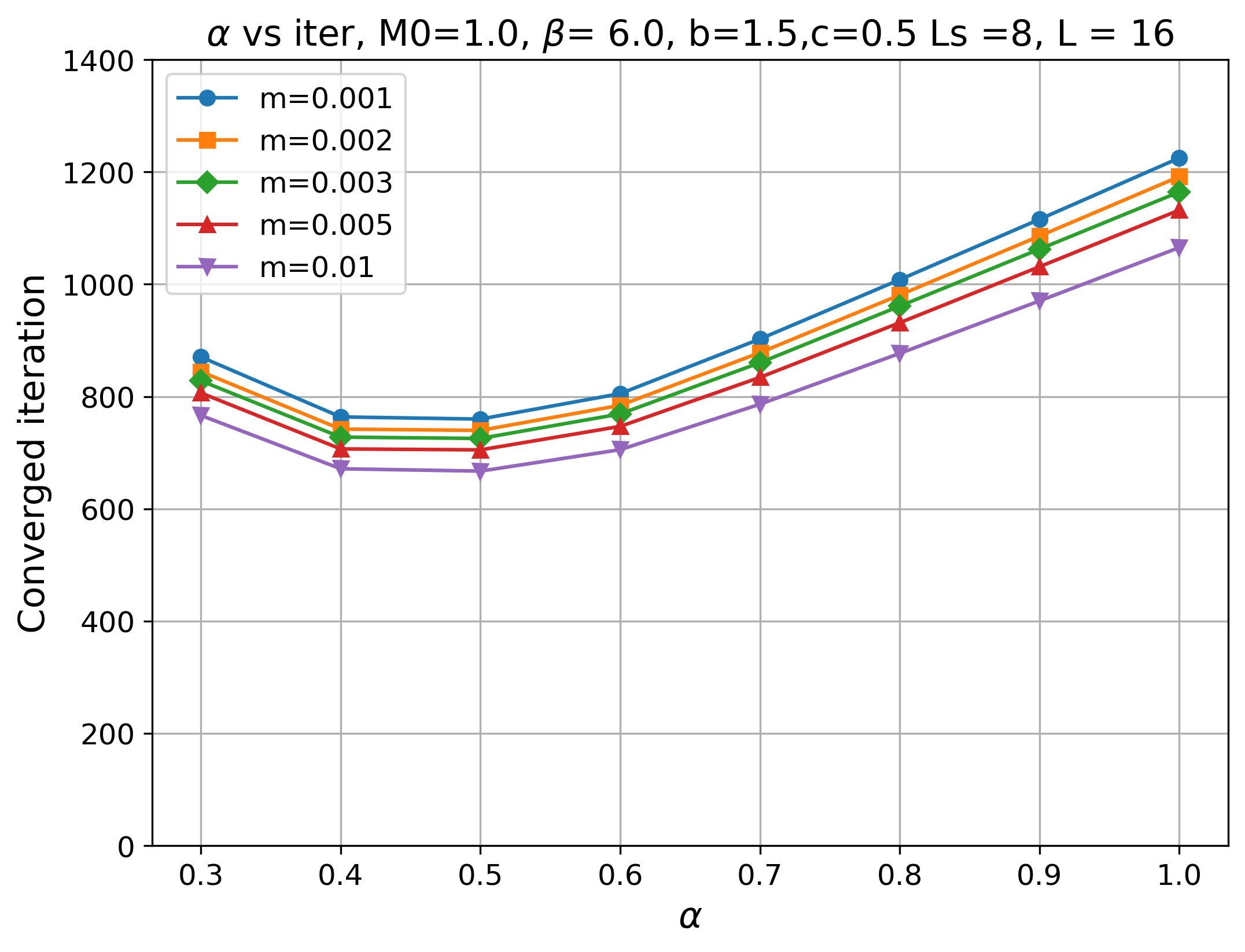}
\includegraphics[width=0.49\textwidth]{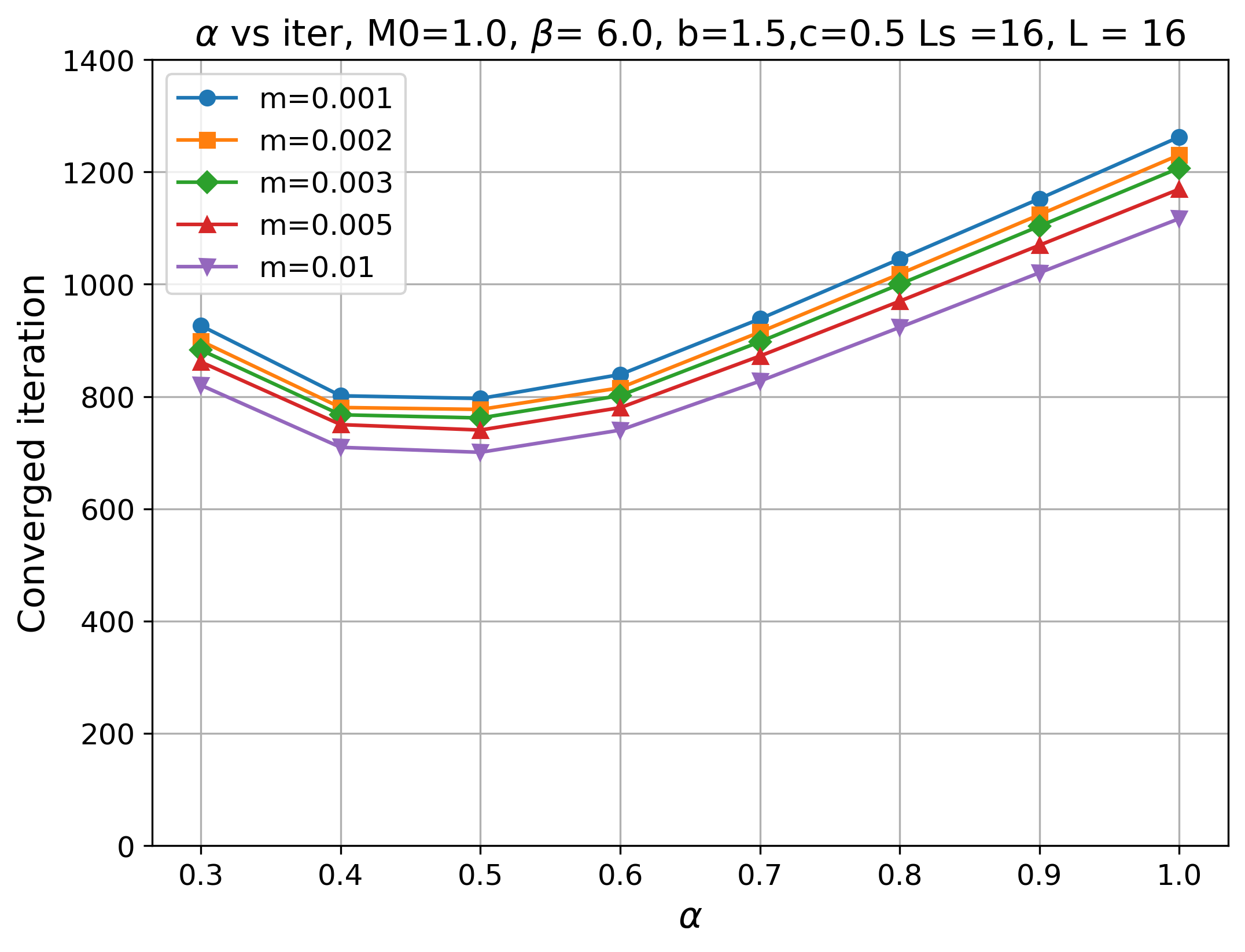}
\includegraphics[width=0.49\textwidth]{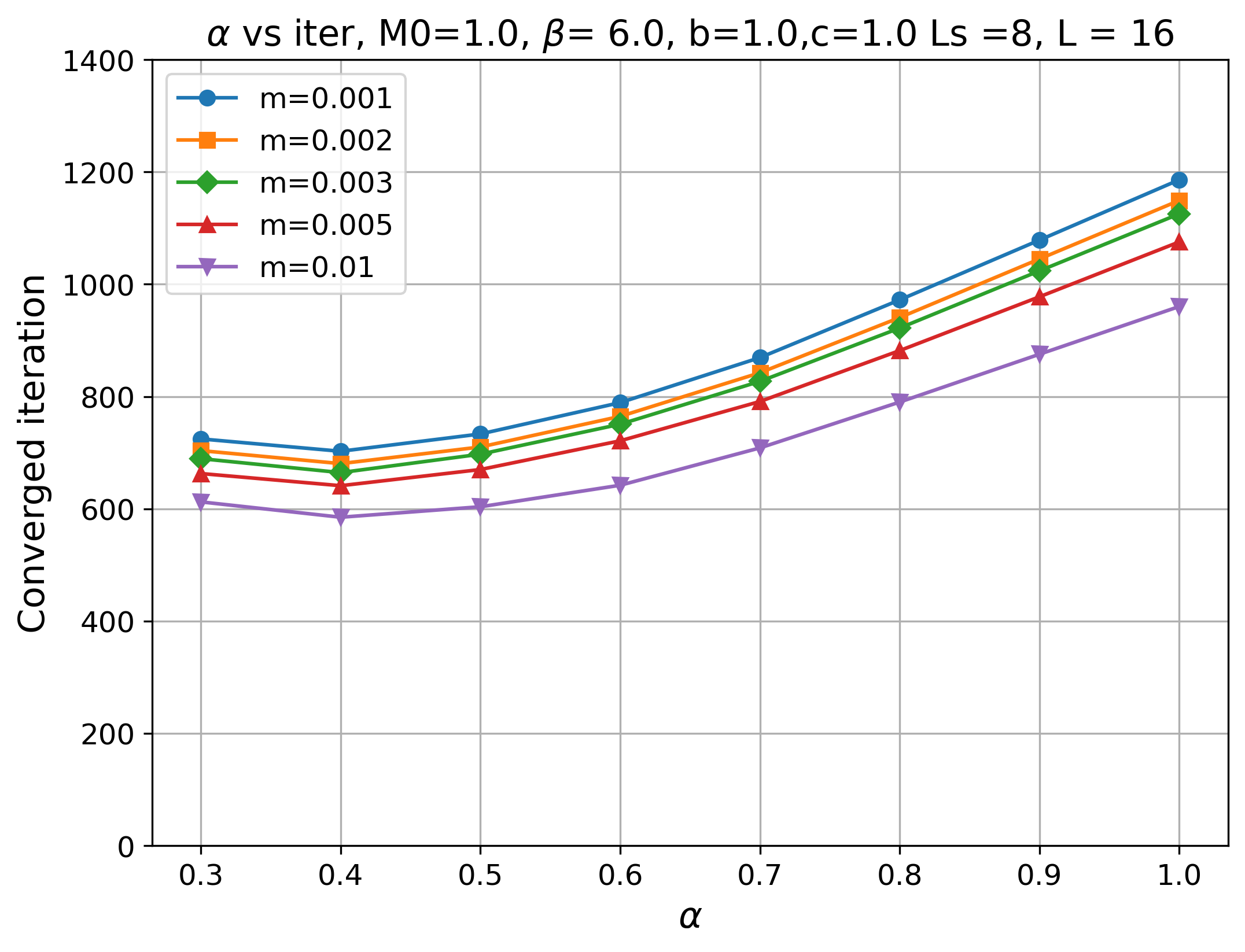}
\includegraphics[width=0.49\textwidth]{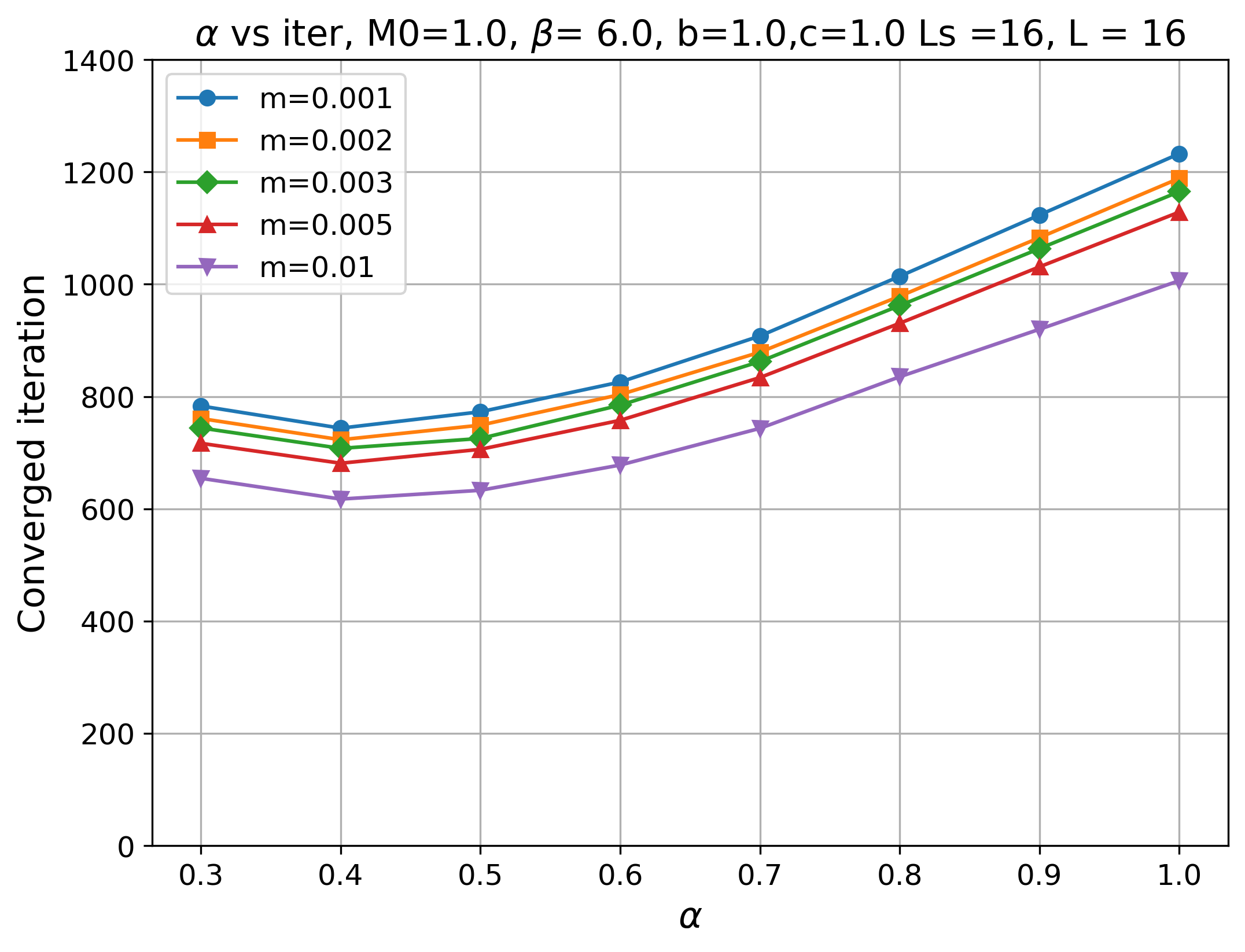}
\caption{
Converged number of CG iteration for $\beta=6.0$ on $(L=16)^4$ lattice,
$M_0 = 1.0$ (with smearing), $(b,c)=(1.5,0.5)$
(top panels) $L_s=8$ (left) and 16 (right)
and $(b,c)=(1.0,1.0)$ (bottom panels).
}
\label{fig:conv_L16_beta6.0_M01.0}
\end{figure}

Next, we observe the effect of the link smearing.
For the domain-wall operator with smeared link, we set $M_0=1.0$. 
The top panels of Figure~\ref{fig:conv_L16_beta6.0_M01.0}
show the result for $b=1.5$, $c=0.5$.
Together with the link smearing, this choice of the parameters corresponds to the setup
adopted by the JLQCD Collaboration, except that the configuration in this work is
in the quenched approximation.
Around $\alpha\simeq 0.5$, the converged number of CG iterations
becomes minimum where 37\% acceleration is achieved for $m=0.001$.
We find that changing $L_s$ does not affect the optimal value of $\alpha$,
and has only little effect on the iteration number for the
convergence.

The bottom panels of Figure~\ref{fig:conv_L16_beta6.0_M01.0}
shows the result for $(b,c)=(1.0,1.0)$.
A similar tendency to the $(b,c)=(1.5, 0.5)$ case is found
while the best value of $\alpha$ becomes around 0.4.
The achieved speed-up is about 40 \% for $m=0.001$.
Also small effect of $L_s$ on the convergence is observed.

\subsubsection{Condition number}
The convergence of the CG algorithm is represented by the
condition number which is defined as the ratio of
the maximum and minimum eigenvalues of the matrix.
We measure the minimum and maximum eigenvalues of $D^\dagger D$
by applying the implicitly restarted Lanczos algorithm 
in the double precision which is implemented in the Bridge++ code.

\begin{figure}[tb]
\centering
\includegraphics[width=0.48\textwidth]{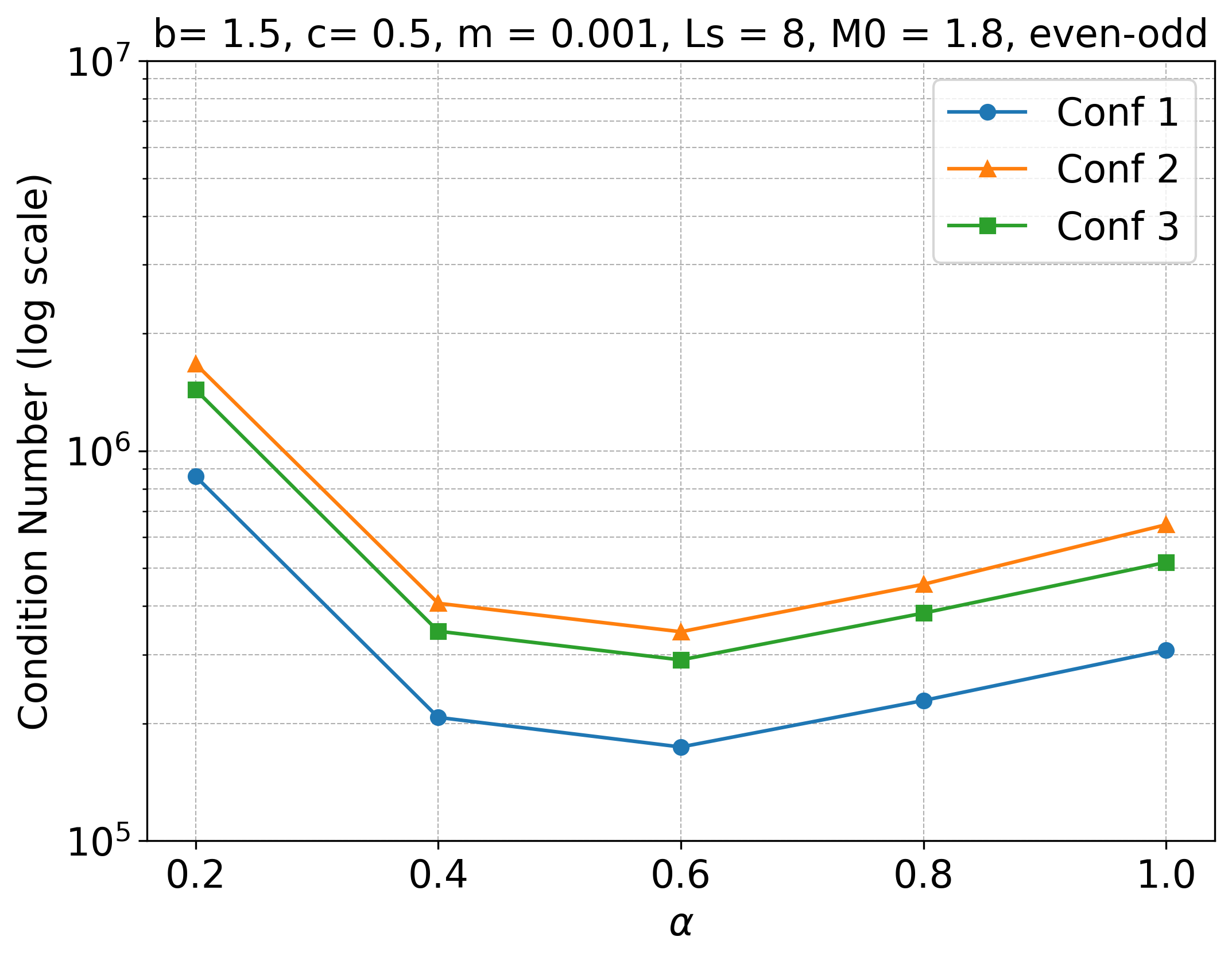}
\includegraphics[width=0.48\textwidth]{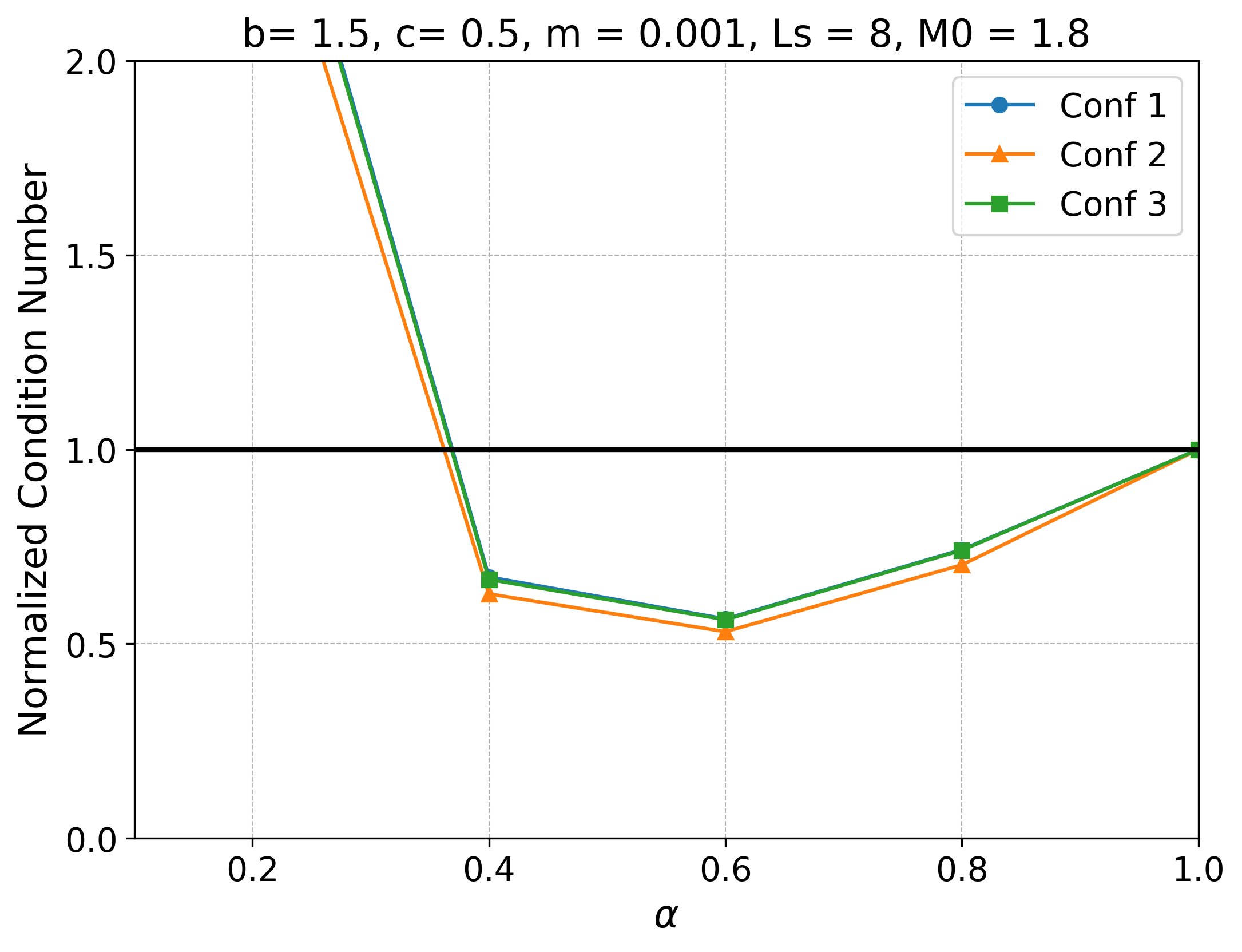}
\includegraphics[width=0.48\textwidth]{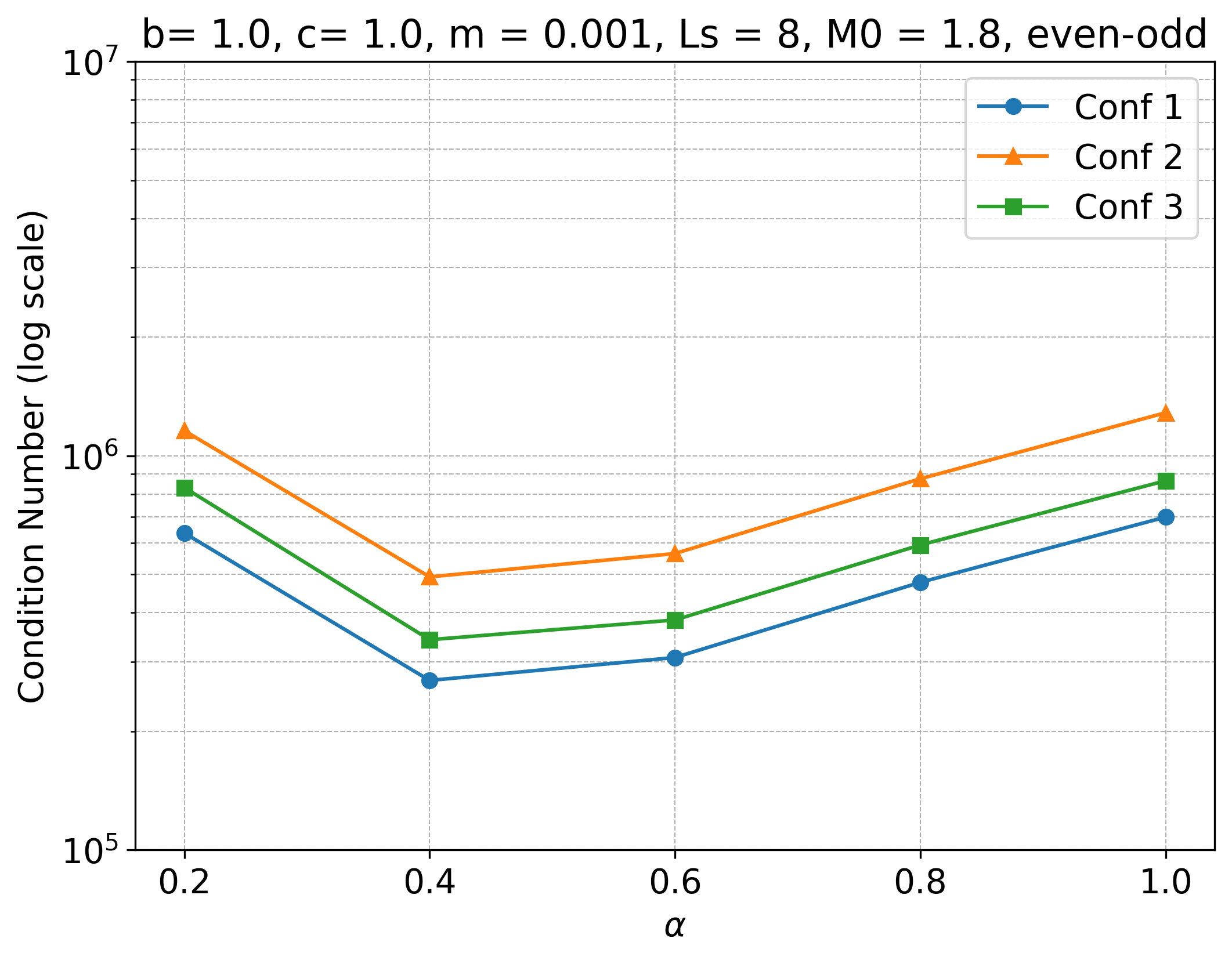}
\includegraphics[width=0.48\textwidth]{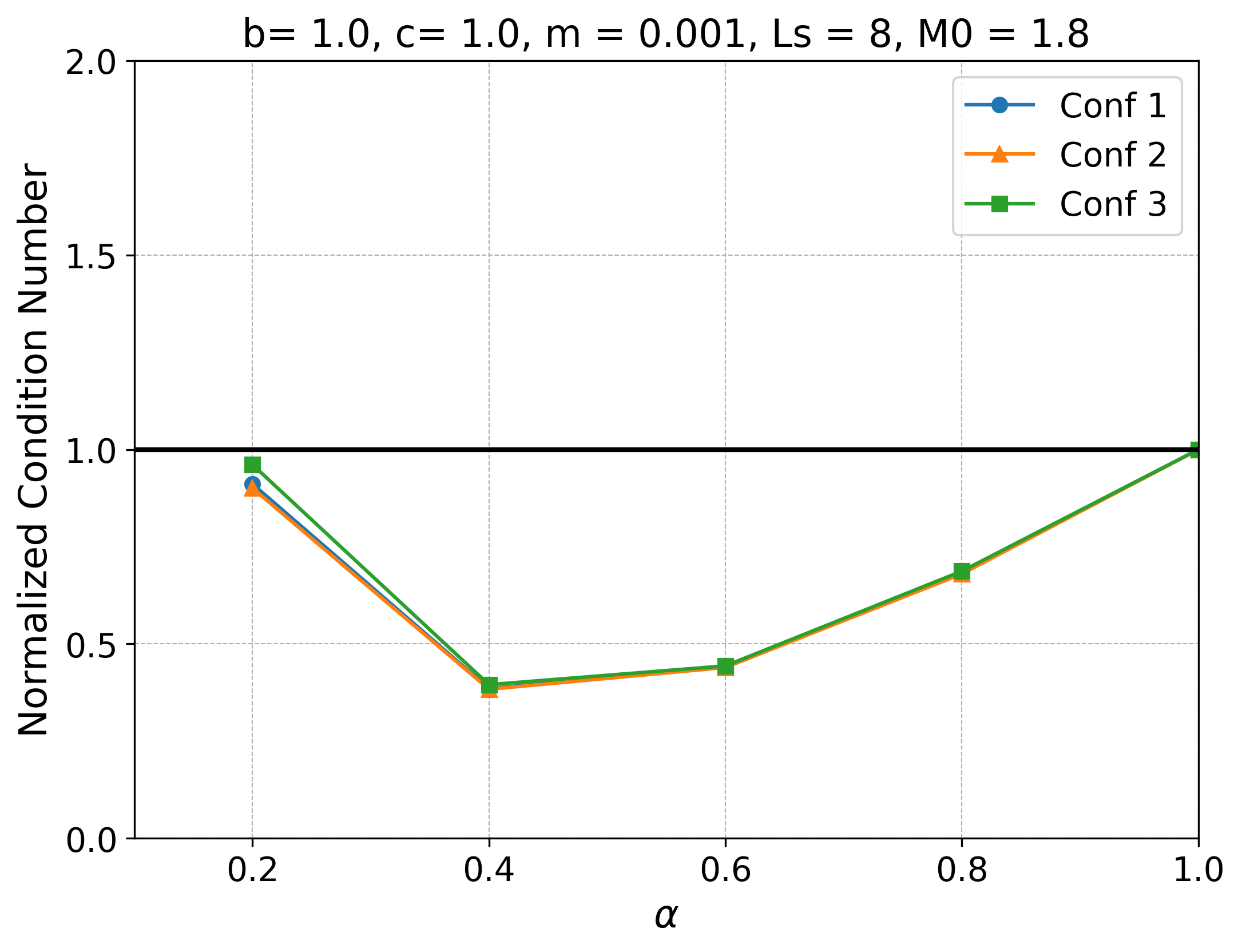}
\caption{
Result of the condition number on the $16^4$ lattice at
$\beta=6.0$ for the domain-wall operator without link smearing
with parameters $M_0=1.8$, $(b,c)=(1.5,0.5)$ (top panels)
and $(1.0,1.0)$ (bottom), $m=0.001$.
The left panels show the observed condition number on
three configurations.
The right panels show the condition numbers normalized
by the values for $\alpha=1$.}
\label{fig:conditionnumber_M518}
\end{figure}

In Figure~\ref{fig:conditionnumber_M518}, the measured
condition number against the value of $\alpha$ is displayed
for the domain-wall operators without the link smearing.
We display the result for $(b,c)=(1.5, 0.5)$ (top panels)
and $(b,c)=(1.0, 1.0)$ (bottom) for $m=0.001$. 
The left panels show the values of the computed condition
number on three configurations.
While the fluctuation of the condition number against
configuration is substantial, as displayed in the right
panels of Figure~\ref{fig:conditionnumber_M518},
normalizing by the values at $\alpha=1$, the effect of
$\alpha$ itself is stable.
The $\alpha$ dependence of the condition number exhibits
similar behavior to the converged number of iterations
shown in Figure~\ref{fig:conv_L16_beta6.0_M01.8},
which is consistent with the theoretical expectation.

\begin{figure}[tb]
\centering
\includegraphics[width=0.48\textwidth]{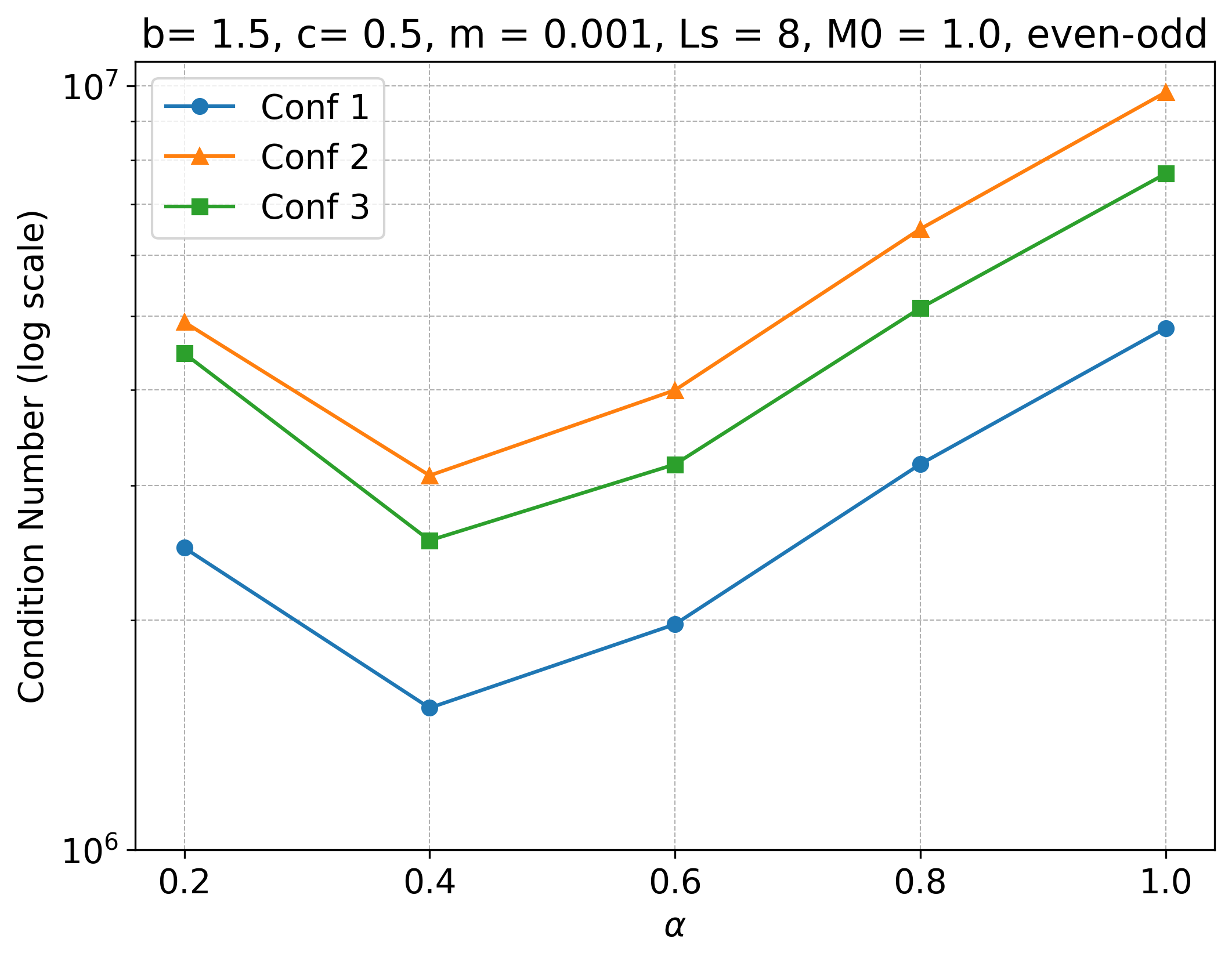}
\includegraphics[width=0.48\textwidth]{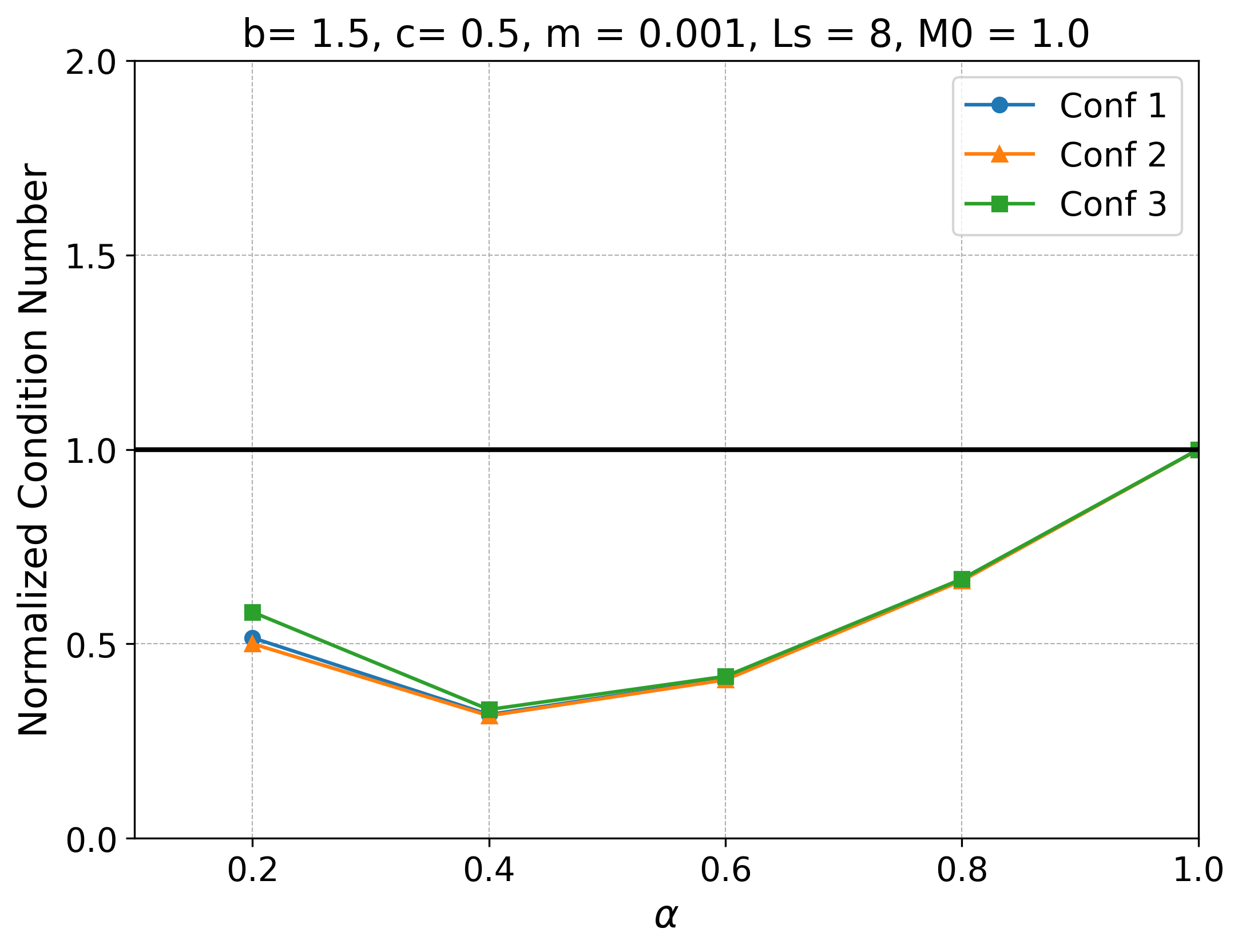}
\includegraphics[width=0.48\textwidth]{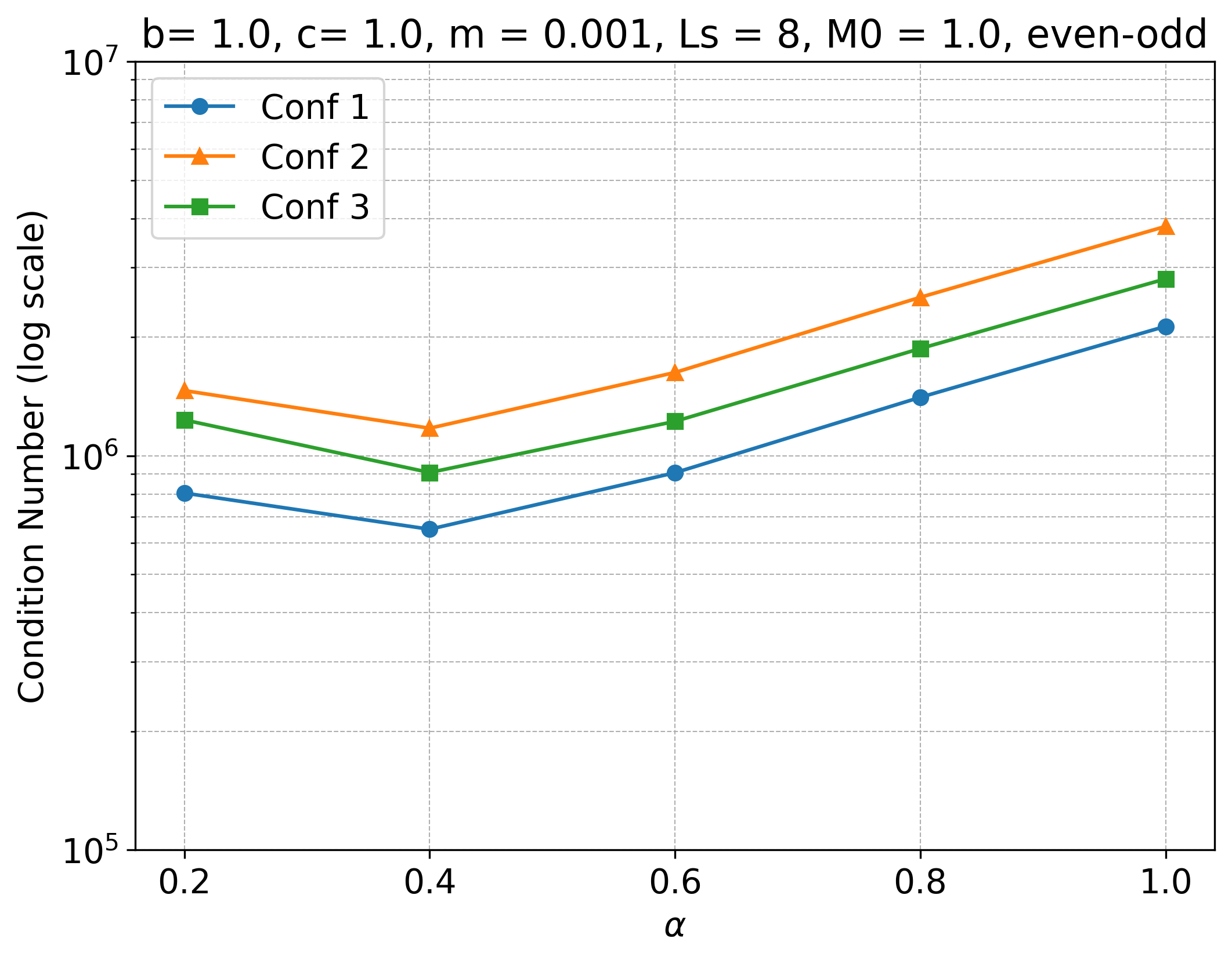}
\includegraphics[width=0.48\textwidth]{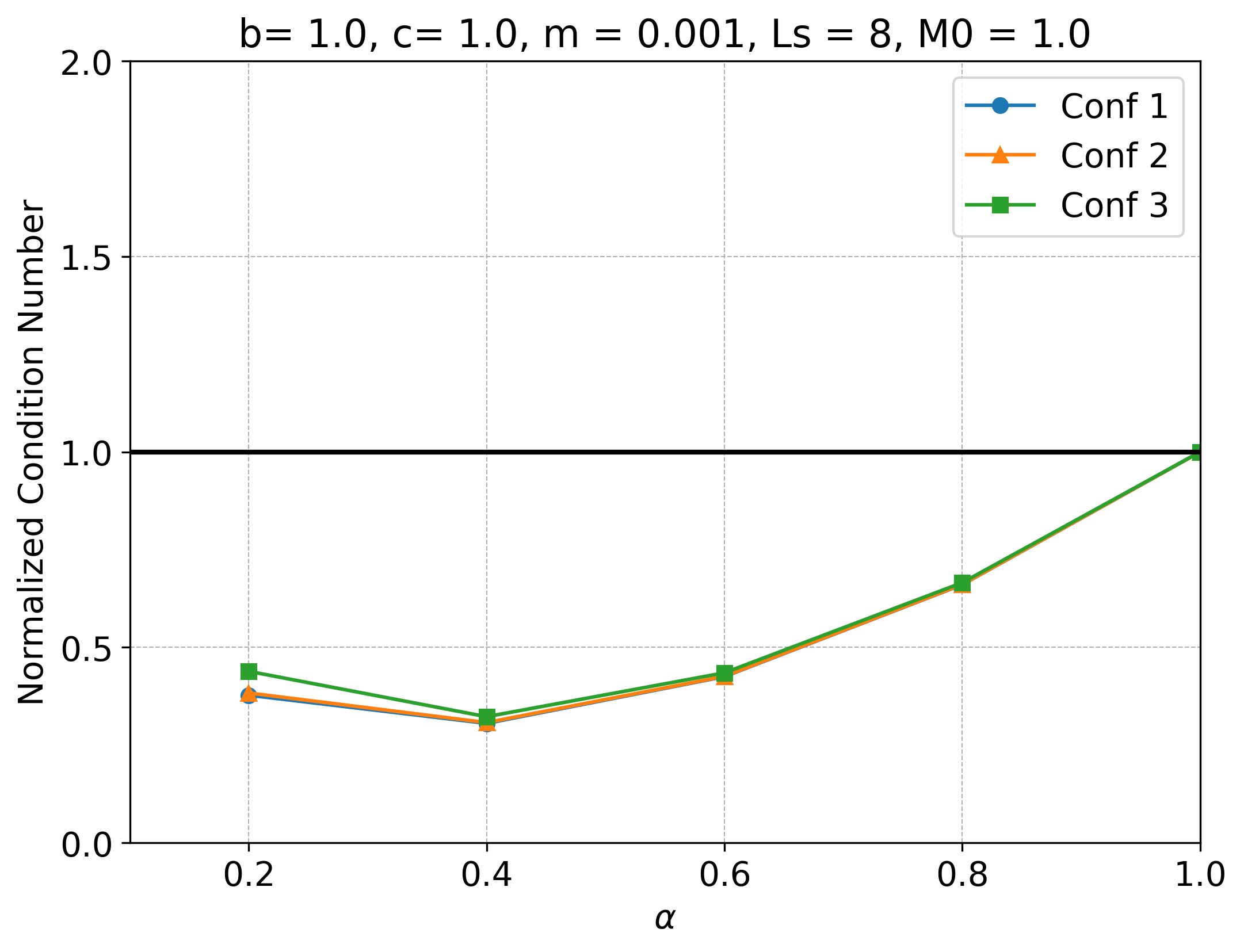}
\caption{
Result of the condition number on the $16^4$ lattice at
$\beta=6.0$ for the domain-wall operator with link smearing
with parameters $M_0=1.0$, $(b,c)=(1.5,0.5)$ (top panels)
and $(1.0,1.0)$ (bottom), $m=0.001$.
The left panels show the observed condition number on
three configurations.
The right panels show the condition numbers normalized
by the values for $\alpha=1$.}
\label{fig:conditionnumber_M510}
\end{figure}

Figure~\ref{fig:conditionnumber_M510} shows the condition numbers
and those normalized by the values at $\alpha=1.0$. 
Similarly to the case without smearing, the behavior of the condition
numbers is consistent with the converged number of CG iterations
in Fig.~\ref{fig:conditionnumber_M510}.
It is also observed that the fluctuation against the configuration is
absorbed by normalizing with the values at $\alpha=1$.

\subsection{Result for $\beta$ = 6.0 on the $32^4$ lattice}

To examine the effect of lattice volume on the convergence
of the CG algorithm, we measure the converged
iteration number against $\alpha$ on a $32^4$ lattice
with configurations generated at the gauge coupling $\beta=6.0$.
This means that the lattice volume is doubled in each direction
compared to the lattice examined in the previous subsection.
On this lattice size, we only show the results for $L_s=8$.

\begin{figure}[tb]
\centering
\includegraphics[width=0.48\textwidth]{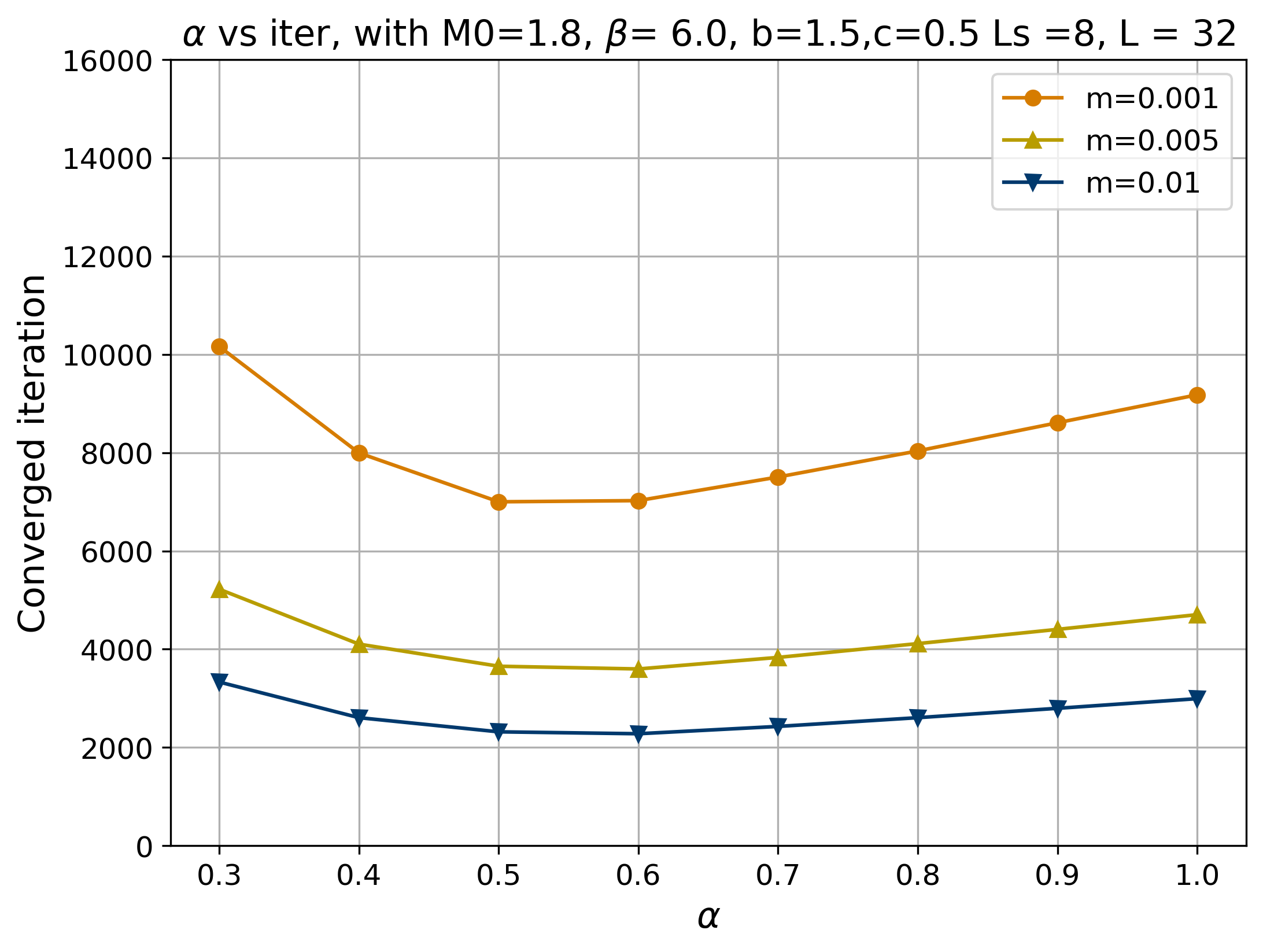}
\includegraphics[width=0.48\textwidth]{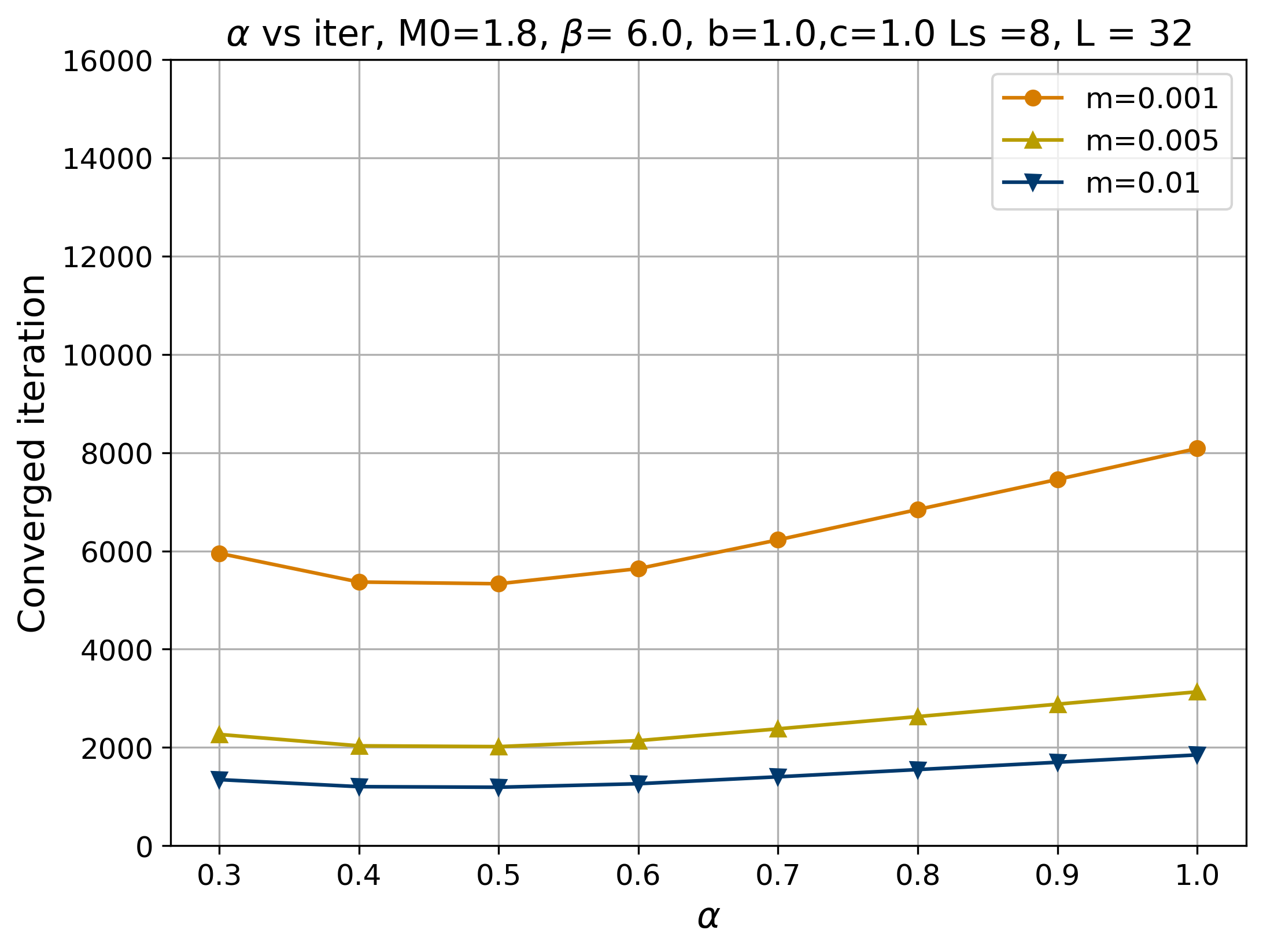}
\includegraphics[width=0.48\textwidth]{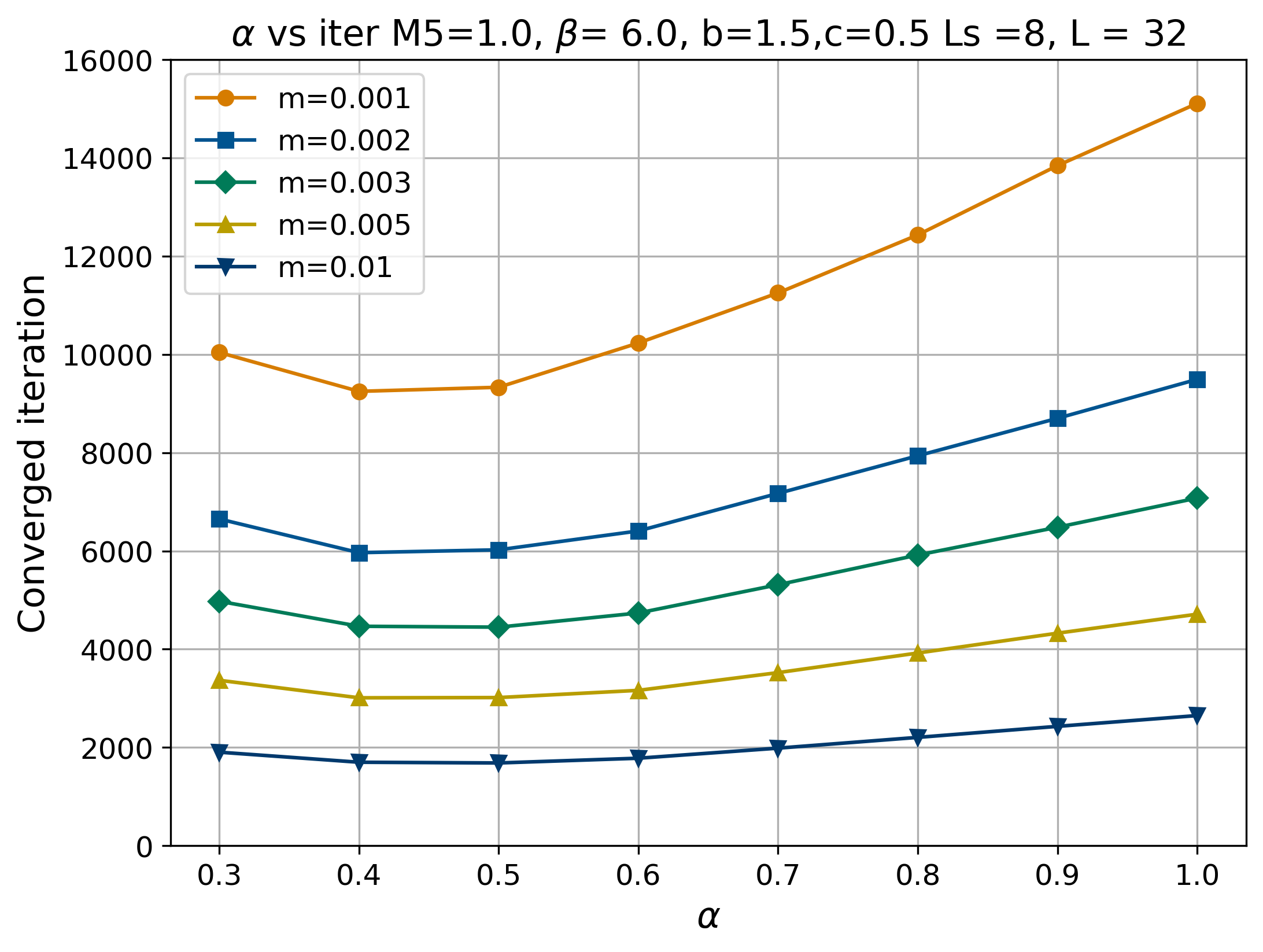}
\includegraphics[width=0.48\textwidth]{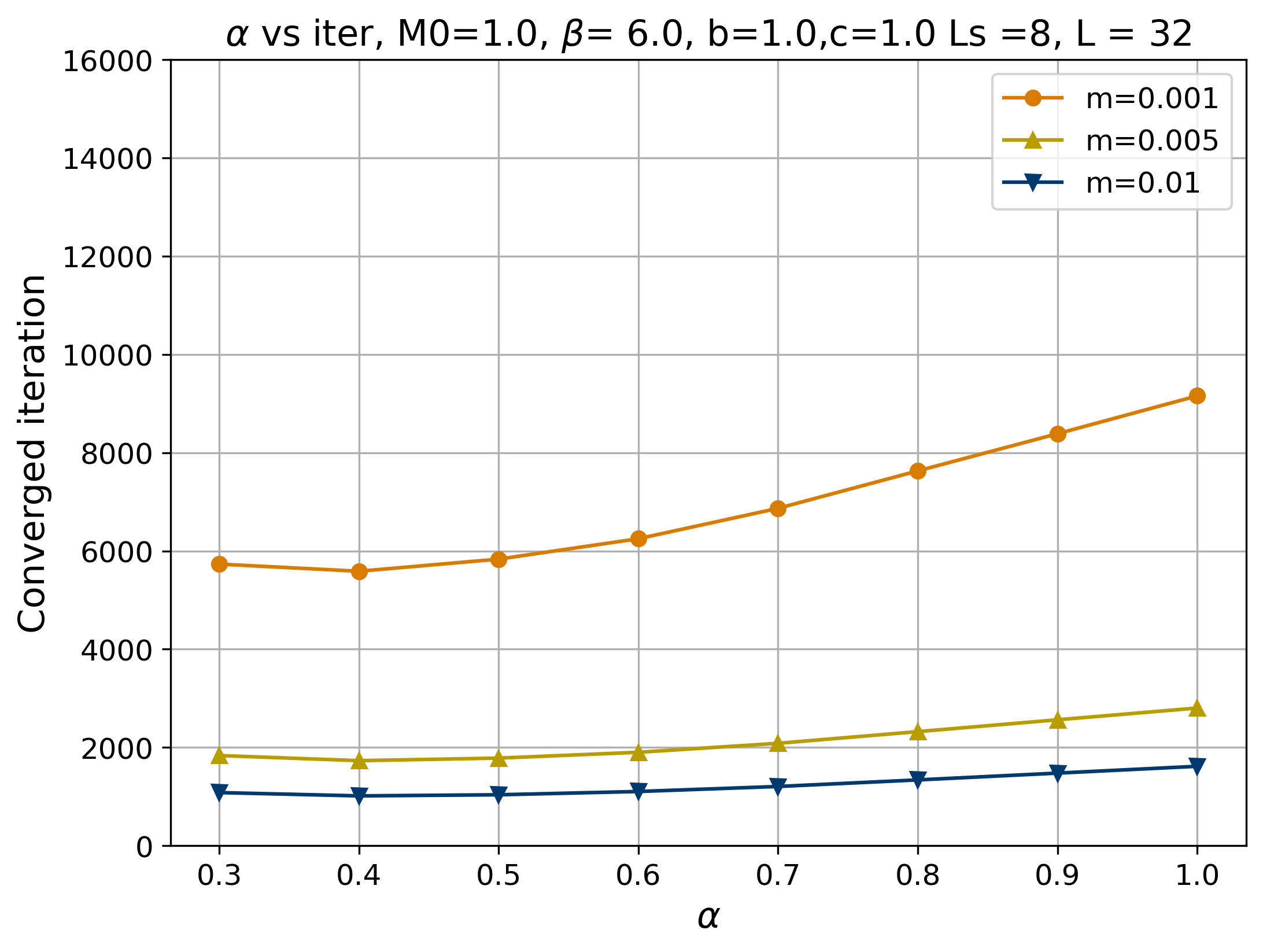}
\caption{
Converged numbers of CG iterations on $32^4$ lattice
at $\beta=6.0$.
Top panels show $M_0 = 1.8$ (without smearing),
$(b,c)=(1.5,0.5)$ (left) and $(b,c)=(1.0,1.0)$ (right).
Bottom panels show $M_0=1.0$ (with smearing),
$(b,c)=(1.5,0.5)$ (left) and $(b,c)=(1.0,1.0)$ (right).
In both cases, $L_s=8$ is used.
}
\label{fig:conv_L32_beta6.0_M01.0_b1.5_c0.5_Ls8}
\end{figure}

Figure~\ref{fig:conv_L32_beta6.0_M01.0_b1.5_c0.5_Ls8} shows
the results of the converged number of CG iterations
for unsmeared and smeared cases with $(b,c)=(1.5,0.5)$
and $(1.0,1.0)$.
The acceleration of CG convergence for $m=0.001$, the severest
one of the measured quark mass, is in the range of
24\% (top-left panel)--40\% (bottom-left).

As a different feature from the results on $16^4$ lattice in
Figs~\ref{fig:conv_L16_beta6.0_M01.8} and
\ref{fig:conv_L16_beta6.0_M01.0}, large dependence on
the quark mass is observed.
This is considered that $16^4$ lattice may be affected by
finite volume effect, since for the present small values of
quark mass, the lattice size of about 1.6 fm is not
sufficiently larger than the pion Compton wavelength.
This should be examined by observing the pion spectrum
by varying the lattice volume.

In the context of this paper, we focus on the effect of
$\alpha$ on the convergence number of iterations.
As Figure~\ref{fig:conv_L32_beta6.0_M01.0_b1.5_c0.5_Ls8} 
shows, the optimal values of $\alpha$ are almost
unchanged from those on the $16^4$ lattice at the same
$\beta$ with similar improving rates.
This result implies that incorporating $\alpha$ in
the domain-wall operator results in improved convergence
of the even-odd preconditioned CG solver almost
independently of lattice volume.

\subsection{Result for $\beta$ = 5.7 on the $16^4$ lattice}

We also examine a $16^4$ lattice with configurations
generated at $\beta=5.7$, which roughly corresponds to
the lattice spacing $a\simeq 0.2$ fm.
This implies that the physical volume is almost the same as that of $32^4$
lattice at $\beta=6.0$.
We adopt the same values of the quark mass $m$ in lattice units as those for
$\beta=6.0$, which means that the mass in physical units is halved correspondingly.

\begin{figure}[tb]
\centering
\includegraphics[width=0.48\textwidth]{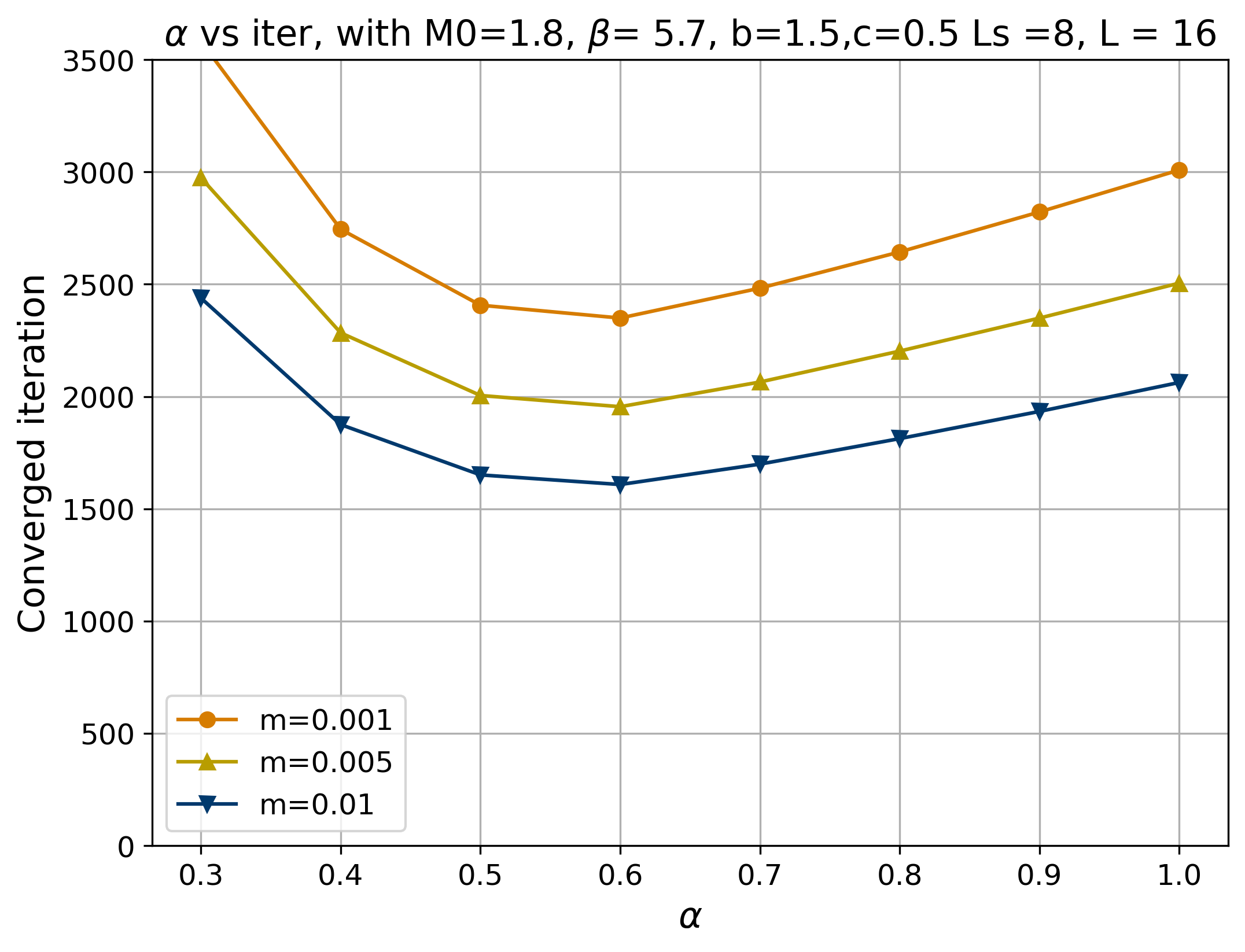}
\includegraphics[width=0.48\textwidth]{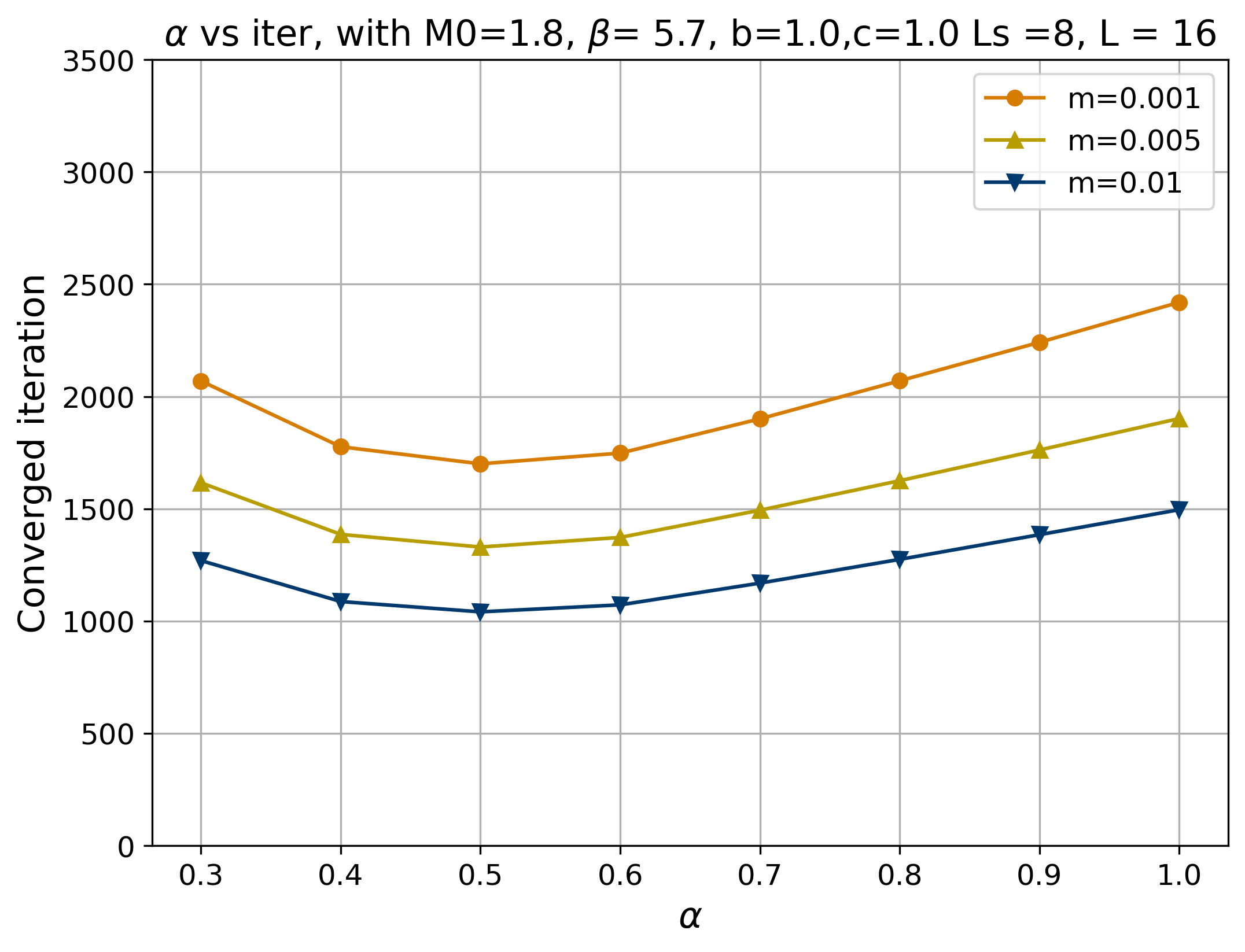}
\includegraphics[width=0.48\textwidth]{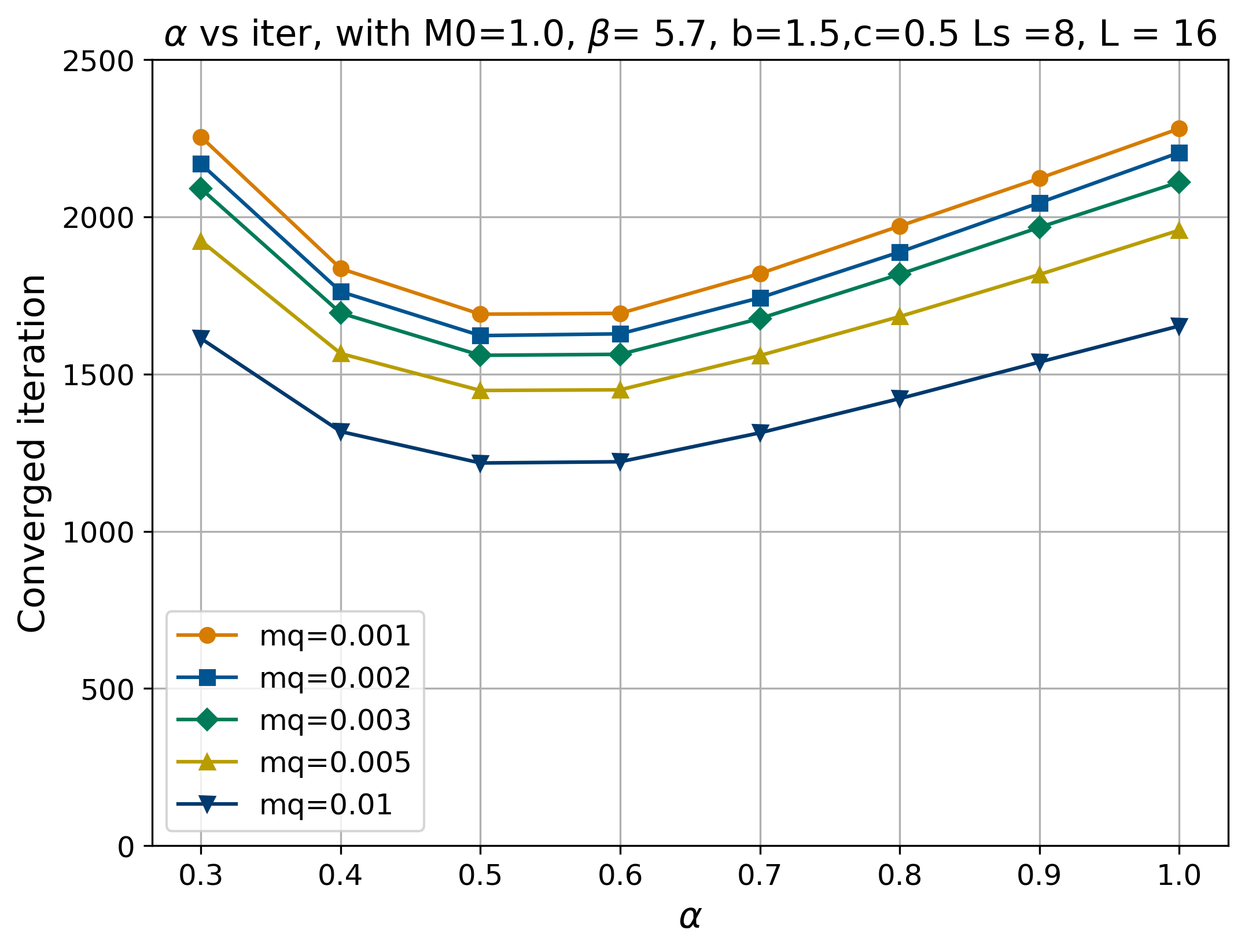}
\includegraphics[width=0.48\textwidth]{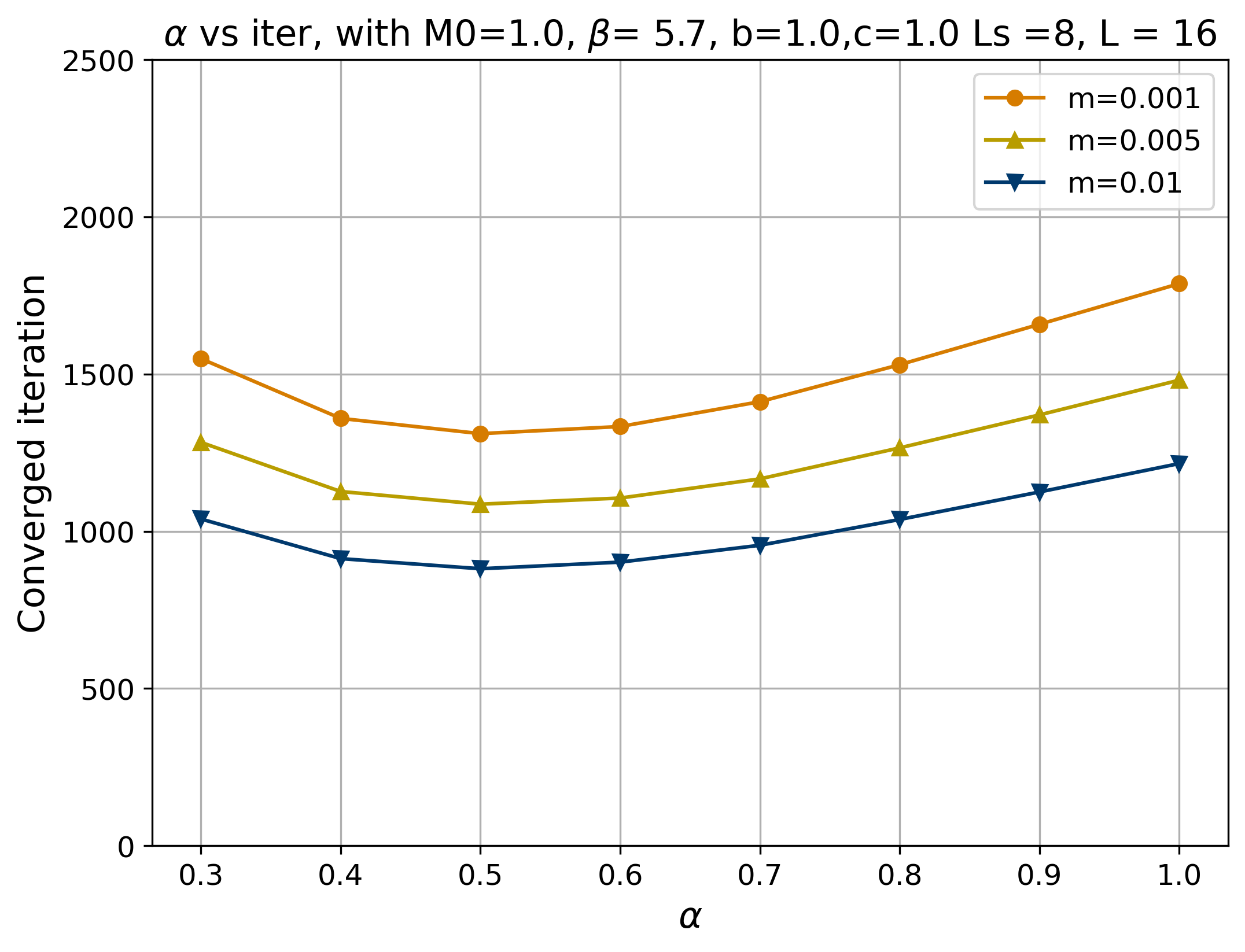}
\caption{
Converged numbers of CG iterations on $16^4$ lattice
at $\beta=5.7$.
Top panels show $M_0 = 1.8$ (without smearing),
$(b,c)=(1.5,0.5)$ (left) and $(b,c)=(1.0,1.0)$ (right).
Bottom panels show $M_0=1.0$ (with smearing),
$(b,c)=(1.5,0.5)$ (left) and $(b,c)=(1.0,1.0)$ (right).
In both cases, $L_s=8$ is used.
}
\label{fig:conv_L16_beta5.7_M01.0_b1.5_c0.5_Ls8}
\end{figure}

Figure~\ref{fig:conv_L16_beta5.7_M01.0_b1.5_c0.5_Ls8} shows that the $\alpha$ improvement effect diminishes as the quark mass becomes larger. 
Similar to the case of $\beta=6.0$, the optimal improvement is observed when $\alpha$ is in the range of $0.4$ to $0.5$. 
The acceleration of CG convergence for $m=0.001$ is in the range of
20\% (top-left panel)--30\% (top-left).

\section{Conclusion and outlook}
\label{sec:Conclusion}

In this paper, we examined the effect of the improved form of
the domain-wall fermion operator on the convergence of the linear equation
solver.
In a wide range of parameters, this form indeed improves the
convergence of the even-odd preconditioned CG solver by about several tens of percent,
in the range of 20-40\% for our smallest value of the quark mass.
The required modification of the simulation code is small and the increase
in the arithmetic operations is negligible compared to the observed gain.
Indeed, in a modern simulation environment where the memory bandwidth or communication
overhead is more relevant to the performance than the arithmetic operations,
this modification has little effect on the simulation time.
Thus the practical use of the improved form of the domain-wall fermion
matrix is highly attractive.

In a future release, the Bridge++ code set will incorporate the
improved form as the standard implementation of the domain-wall
fermion matrix.

\subsection*{Acknowledgment}

We thank the members of the Bridge++ project for useful discussions.
The code development and measurement were performed on
the Cygnus and Pegasus systems at University of Tsukuba and
Wisteria/BDEC-01 at University of Tokyo through the Multidisciplinary
Cooperative Research Program in Center for Computational Sciences,
University of Tsukuba.
The code development was also done on the Supercomputer Fugaku
(through Usability Research ra000001)
at RIKEN Center for Computational Science.
This work is supported by JSPS KAKENHI (JP20K03961, JP22H00138, JP23K22495),
MEXT as “Program for Promoting Researches on the Supercomputer Fugaku”
(Simulation for basic science: approaching the new quantum era,
PMXP1020230411)
and Joint Institute for Computational Fundamental Science (JICFuS).


\bibliographystyle{splncs04}
\bibliography{iccsa2025dwf}

\end{document}